\documentclass{article}
\usepackage{graphicx} 
\usepackage{amsmath}
\usepackage{latexsym}
\usepackage{xcolor}

\def\hybrid{
        \topmargin -20pt
        \oddsidemargin 0pt
        \headheight 0pt \headsep 0pt
        \textwidth 6.25in 
        \textheight 9.5in 
        \marginparwidth .875in
        \parskip 5pt plus 1pt \jot = 1.5ex}

\hybrid

\begin{document}
\begin{titlepage}
\rightline{}
\begin{center}
\vskip 1.5cm
{\Large \bf{Generalized Bergshoeff-de Roo identification in the Supergravity frame}}
\vskip 1.7cm

{\large\bf {Walter H. Baron$^1$, Fabian A. Portilla$^{2}$}}
\vskip 1cm

$^1$ {\it  Instituto de F\'isica de La Plata}, (CONICET-UNLP)\\
{\it Departamento de Fisica, Universidad Nacional de La Plata. }\\
{\it C. 115 s/n, B1900, La Plata, Argentina}\\
wbaron@fisica.unlp.edu.ar\\
 
\vskip .3cm

$^2$ {\it Academia Tomas Cipriano de Mosquera}\\
{\it Departamento de Física}\\
{\it Transversal 9 \#61-56A Popayán, Colombia}\\
fabianportilla@unicauca.edu.co

\vskip .3cm

\vskip .1cm

\vskip .4cm

\vskip .4cm

\end{center}
\bigskip\bigskip

\begin{center} 
\end{center} 
\begin{quote}  
The effective action of string theory has a highly constrained structure due to the numerous symmetries of the underlying theory. Recently, a method known as the Generalized Bergshoeff-de Roo identification was developed, which exploits diffeomorphism, Lorentz and gauge invariance, along with T-duality, to derive higher-derivative couplings in the effective action within the framework of Double Field Theory. This method has been revisited and further refined to generate couplings directly within the standard Supergravity framework.    
\end{quote} 
\vfill
\setcounter{footnote}{0}
\end{titlepage}

\section{Introduction and summary of results}
Symmetries play an essential role in the formulation of physical theories. In some cases, their physical interpretation is transparent; in others, they are of a hidden nature, emerging as vestiges of a symmetry inherited from a higher-energy parent theory. Regardless of their origin, symmetries serve as powerful tools that can completely or partially fix the couplings of the theory. A paradigmatic example of the latter is T-duality in low-energy effective string theories.

There are, however, instances where the origin of the symmetry is less clear. This is precisely the case for a particular symmetry realized in $\mathcal{N}=1$ supergravity coupled to gauge fields~\cite{Bergshoeff:1988nn}.

Bergshoeff and de Roo realized that the torsionful spin connection and the gravitino curvature transform under supersymmetry exactly as a vector multiplet:
\begin{eqnarray}
&&\;\;\;\;\;\;\delta A_{m}=\frac12\bar\varepsilon \Gamma_{m}\chi\;,
\;\;\;\;\;\;\;\;\;
\delta\chi=\frac14 \Gamma^{mn} \hat F_{mn} \varepsilon +\frac{\sqrt2}{2} \left(\vphantom{\frac12}\varepsilon \bar \chi\lambda 
-\chi\bar\varepsilon\lambda + \Gamma^a \lambda \bar\chi \Gamma_a\varepsilon
\right)\;,
\cr&&\cr
&& \delta \omega^{(-)}_{m}{}^{ab}=\frac12\bar\varepsilon \Gamma_{m}\psi^{ab}\;,\;\;\;
\delta\psi^{ab}=-\frac14 \Gamma^{mn} \hat R^{(-)}_{mn}{}^{ab} \varepsilon +\frac{\sqrt2}{2} \left(\vphantom{\frac12}\varepsilon \bar \psi^{ab}\lambda 
-\psi^{ab}\bar\varepsilon\lambda + \Gamma^a \lambda \bar\psi^{ab}\Gamma_a\varepsilon\right)\;.
\;\;\;\label{SUSY}
\end{eqnarray}

This symmetry under the exchange of gauge degrees of freedom and (composite) gravitational degrees of freedom suggests that the action itself should reflect it. 

A remarkable fact about this symmetry is that it connects fields with different orders in derivatives and, in principle, allows access to higher-derivative corrections in the action. In particular, the kinetic terms $|F|^2$ in the action suggest the inclusion of quadratic $R^2$ terms, which had already been predicted by other computations.

A limitation of this approach, however, is that the interactions induced in the action modify the supersymmetry transformations at higher derivative orders. Since such modifications explicitly break the original symmetry (\ref{SUSY}), this method cannot be employed straightforwardly to access higher orders. Results beyond the four-derivative couplings were derived in \cite{Bergshoeff:1988nn} using complementary techniques.

\bigskip

{\bf{{The generalized Bergshoeff-de Roo identification}}}
\medskip

A generalization of this approach was proposed in \cite{Baron:2018lve} within the framework of the so-called Double Field Theory (DFT) \cite{Siegel,DFT} (see \cite{Reviews} for reviews). 

This theory, in its simplest version describing the NS-NS sector, is a field theory in which T-duality is promoted to a manifest symmetry. This is achieved, on the one hand, by arranging the fields into multiplets of $O(D,D)$ with $D=10$ the spacetime dimension and, on the other hand, by doubling the number of coordinates, so that the derivatives of the fields also fit into multiplets of $O(D,D)$. 

Consistency of the theory requires imposing certain constraints on the derivatives of the fields and the parameters of the theory. The most conservative solution (which we assume here) restricts the dependence of the fields to the usual 10 physical dimensions, and DFT reduces to generalized geometry with local symmetry transformations parametrized by $\xi^{M}=(\xi_m,\xi^{m})$ and $\Gamma^{AB}=(\Gamma^{\underline{ab}}, \Gamma^{\overline{ab}})$, with $\xi^m$ and $\xi_m$ representing the parameters of diffeomorphisms $\delta x^m= \xi^m$ and gauge transformations of the Kalb-Ramond field $\delta b_{mn} =  2\partial_{[m}\xi_{n]}$, and $\Gamma^{AB}$ the parameters of an extended Lorentz group. 

A gauged DFT was constructed in \cite{Hohm:2011nu} (see also \cite{Jeon:2011sq}). There, the gauge fields, the vielbein, and the 2-form are arranged into multiplets of $O(D,D+k)$ in a generalized vielbein ${\cal E}_{\cal M}{}^{\cal A}$, and the local transformations are parametrized in terms of $\xi^{\cal M}=(\xi_m,\xi^m,\xi^{\tilde\mu})$, with $\xi^{\tilde\mu}$ describing the parameters of the gauge transformations of the field $A_{m}{}^{\tilde\mu}$, and $\Gamma^{\cal A B}$ describing the parameters of a further extended Lorentz group, with ${\cal A}=\{\underline{a},\overline{\cal A}\}$ and $\overline{\cal A}=\{\overline{a},\overline{\alpha}\}$. 
\begin{eqnarray}
\delta {\cal E}_{\cal M}{}^{\cal A}=\xi^{\cal N}\partial_{\cal N}{\cal E}_{\cal M}{}^{\cal A} 
+  \left( \partial_{\cal M}\xi^{\cal N} - \partial^{\cal N}\xi_{\cal M} \right) {\cal E}_{\cal N}{}^{\cal A} 
+{\hat f}_{\cal M N}{}^{\cal P} \xi^{\cal N} {\cal E}_{\cal P}{}^{\cal A} 
+ {\cal E}_{\cal M}{}^{\cal B} \Gamma_{\cal B}{}^{\cal A} \;. \label{TransformationExtended}
\end{eqnarray}

Here $\hat f_{\cal MN}{}^{\cal P}= g \ f_{\cal MN}{}^{\cal P}$, $g$ is the gauge coupling constant and $f$ represents the structure constants of the gauge algebra, which for us will be non-zero when $\{{\cal M,N,P}\}=\{\tilde\mu,\tilde\nu,\tilde\rho\}$, and zero otherwise. 

In \cite{Baron:2018lve}, an extension of the identification found in \cite{Bergshoeff:1988nn} was obtained within a T-duality covariant framework and with some conceptual differences with respect to the original approach. In the next section we will be much more explicit; here we content ourselves with a general description:

\begin{enumerate}
    \item Now the gravitational and gauge degrees of freedom are contained in a single generalized tensor. In particular, the gauge field is contained within the generalized vielbein, while the torsionful spin connection is contained within the so-called generalized flux ${\cal F}_{\cal ABC}$
\begin{eqnarray}
{\cal F}_{\cal A B C}&=&
3\;{\cal D}_{[\cal A}{\cal E}^{\cal N}{}_{\cal B} {\cal E}^{\cal P}{}_{\cal C]} \; \eta_{\cal N P}
+ {\hat f}_{\cal M N P} {\cal E}^{\cal M}{}_{\cal A} {\cal E}^{\cal N}{}_{\cal B} {\cal E}^{\cal P}{}_{\cal C} \ , \label{ExtendedFluxes}
\end{eqnarray}
here ${\cal D}_{\cal A}\,= {\cal E}^{\cal M}{}_{\cal A} \partial_{\cal M}$ is a flattened partial derivative and $\eta$ represents the invariant metric of $O(D,D+K)$ (see Eq. (\ref{eta})).

Then the extension of the identification found in \cite{Bergshoeff:1988nn} will identify certain components of the generalized vielbein and the generalized flux, hence derivatives of the same object. 

\item The identification can be found even by restricting to the bosonic sector, by analyzing the local transformations (\ref{TransformationExtended}), i.e. without the need to analyze the supersymmetry transformations.\footnote{These are relevant if one wishes to extend the formalism to the fermionic sector. See, for example, section 4 in \cite{Baron:2018lve} or reference \cite{Lescano:2021guc}.} 

Indeed, the transformation (\ref{TransformationExtended}) induces 
\begin{eqnarray}
\delta {\cal F}_{\cal A B C}=\xi^{\cal M}\partial_{\cal M}{\cal F}_{\cal A B C} \;
+ 3\;\left(\vphantom{\frac12}\Gamma^{D}{}_{[\cal A}\; {\cal F}_{\cal B C] D}- {\cal D}_{[\cal A} \Gamma_{\cal BC]}\right)\, \;. \label{TransfFluxes}
\end{eqnarray}

The generalized Bergshoeff-de Roo (gBdR) identification arises by comparing the transformations of ${\cal E}_{\mu}{}^{\underline{a}}$ and ${\cal F}_{\underline{a}\overline{BC}}$, which contain in their parametrization the degrees of freedom of the gauge fields $A_{m}{}^{\overline\alpha}$ and the torsionful spin connection, respectively. To make the identification more transparent, we use the generators of the gauge algebra $(t^{\tilde{\mu}})_{\overline{\cal BC}}$ and introduce
\begin{eqnarray}
{\cal E}_{\underline a \overline {\cal B C}} = -g  \, {\cal E}_{\tilde\mu \underline a} \, (t^{\tilde\mu})_{\overline {\cal B C}} \ , \ \ \ \xi_{\overline {\cal {BC}}} = - g  \, \xi_{\tilde \mu} \, (t^{\tilde\mu})_{\overline {\cal B C}}\ , \label{IndexRelation}
\end{eqnarray}
One readily verifies
\begin{eqnarray}
 \delta {\cal E}_{\underline a \overline{\cal B C}}&=& \widehat {\cal L}_\xi {\cal E}_{\underline a \overline{\cal B C}} - {\cal D}_{\underline a} \xi_{\overline{\cal B C}} +  2 {\cal E}_{\underline a \overline{\cal D} [\overline{\cal C}}\, \xi^{\overline{\cal D}}{}_{\overline{\cal B}]} +  {\cal E}_{\underline d \overline{\cal B C}}\, \Gamma^{\underline{d}}{}_{\underline a}\; ,\cr
 \delta {\cal F}_{\underline a \overline {\cal B C}} &=& \widehat {\cal L}_\xi {\cal F}_{\underline a \overline {\cal B C}} - {\cal D}_{\underline a} \Gamma_{\overline{\cal B C}} + 2 {\cal F}_{\underline a  \overline {\cal D} [\overline{\cal C}} \Gamma^{\overline{\cal D}}{}_{\overline{\cal B}]} + {\cal F}_{\underline d  \overline {\cal B C}} \Gamma^{\underline d}{}_{\underline {a}} \ . \label{TransformationAH}
\end{eqnarray}
which generalizes the symmetry between gauge and gravitational degrees of freedom found in \cite{Bergshoeff:1988nn}.

\item This is not merely a symmetry between gauge and gravitational degrees of freedom, but rather their outright identification. Indeed, the formulation of~\cite{Hohm:2011nu} is valid, at least classically, for groups far more general than the gauge groups of the heterotic string.\footnote{Anomaly cancellation ultimately restricts the physical gauge groups to $SO(32)$ and $E_8\times E_8$.} Consequently, the gauge group that appears in this framework is interpreted as an auxiliary entity, distinct from the physical gauge group of the heterotic string. Concretely, this amounts to postulating the identifications
\begin{eqnarray}
\xi_{\overline{BC}}=\Gamma_{\overline{BC}} \qquad\text{and}\qquad 
\mathcal{F}_{\underline a \overline{\mathcal{BC}}} = \mathcal{E}_{\underline a\overline{\mathcal{BC}}} .
\end{eqnarray}

\item Interestingly, this identification turns out to be exact, i.e. if certain components of $\cal F_{\cal ABC}$ are identified with $\cal E_{\cal M}{}^{\cal A}$ and subsequently a generalized flux is computed from the latter, its transformation will not be modified. Thus, the limitations found in the original formulation are overcome, where after performing the first correction the identification broke down \cite{Bergshoeff:1988nn}. Although the components of the vielbein at the lowest order transform covariantly under the Lorentz group, the gauge transformations induce, after the identification $\xi^{\tilde\mu}\to \Gamma^{\overline{AB}}$, Green-Schwarz type transformations on the fields. 

\item In this summary, and throughout this work, we only focus on an identification that reproduces the heterotic case, but this mechanism was extended in \cite{Baron:2020xel} also to more general cases, which include a whole family of T-duality covariant theories described by two parameters $(a,b)$ found in \cite{Marques:2015vua}. Of these, the case $a=0, b\neq0$ identifies with the heterotic string, the case $a=b$ with the bosonic string, the case $a=-b$ with the so-called HSZ theory \cite{Hohm:2013jaa}, and the case $a=b=0$ with type II, which is consistent with the fact that the latter do not contain ${\cal O}(\alpha')$ corrections to the action. However, it should be clarified that the infinite tower of couplings reproduced by this technique, although by construction invariant under diffeomorphisms, Lorentz, gauge, and T-duality to all orders in derivatives, does not reproduce all the interactions of the effective theory. In particular, it cannot reproduce the interactions proportional to $\zeta(3)$ that appear at order $\alpha'{}^3$. Whether these can or cannot be reproduced by an improvement of this technique by introducing a new parameter is not yet clear.   

\end{enumerate}

 In a series of works \cite{Baron:2018lve, Baron:2020xel, Lescano:2021guc, Hronek:2021nqk}, a parametrization of ${\cal E}_{\cal M}{}^{\cal A}$ was used in terms of the generalized vielbein of ungauged DFT, which is why this identification, when implemented iteratively, yielded derivative corrections to the DFT action, as well as corrections to the generalized Green-Schwarz transformations. The contact with the corrections to the supergravity actions became very involved, requiring increasingly complicated field redefinitions (see \cite{Baron:2021yqm} for all-order field redefinition of the vielbein and the dilaton fields), as the number of derivatives increased, to rewrite the T-duality covariant multiplets in terms of the usual supergravity multiplets. 
\medskip

In this work we propose a different parametrization of the generalized vielbein, so that the output of the iterative procedure is already written in terms of supergravity multiplets. We will develop the details in the following sections. For now, we present here, as a summary, the main results. The infinite tower of corrections to the action is contained in\footnote{In order to connect our results with those of \cite{Bergshoeff:1988nn} we need to make a field redefinition such that $\hat{H}\to-\hat{H}$, hence $\omega^{(+)}\to\omega^{(-)}$ and $R^{(+)}\to R^{(-)}$.} 
\begin{eqnarray}
{\cal L}= \sqrt{-\hat{g}}\ e^{-2\hat{\phi}}  \left( \hat{R} - 4\ (\nabla \hat\phi)^2 + 4\  \Box\hat\phi -\frac{1}{12} \hat{H}_{mnp} \hat{H}^{mnp} 
-\frac{1}{4}\frac{1}{g^2 X_R} 
{\cal R}^{(+)}_{mn}{}^{\overline{\cal CD}} \ 
{\cal R}^{(+)}{}^{mn}{}_{\overline{\cal CD}}
\right) \; . \label{FullAction}
\end{eqnarray}
Here, $\hat g,\hat \phi$ and $\hat H$ are covariant tensors under diffeomorphisms, Lorentz and gauge transformations. This action contains explicit higher-derivative couplings in terms of a generalized Riemann tensor ${\cal R}^{(+)}_{mn}{}^{\overline{\cal CD}}=\hat{e}^{m}{}_{\underline{a}}\ \hat{e}^{n}{}_{\underline{b}}\ {\cal R}^{(+)}_{\underline{ab}}{}^{\overline{\cal CD}}$
\begin{eqnarray}
{\cal R}^{(+)}_{\underline{ab}}{}^{\overline{\cal CD}} =
 2\sqrt{2}\left( 
D_{[\underline{a}}{\cal F}_{\underline{b}]}{}^{\overline{\cal CD}} 
+ \omega_{[\underline{ab}]}{}^{\underline{e}} {\cal F}_{\underline{e}}{}^{\overline{\cal CD}} 
- \sqrt{2} \ {\cal F}_{[\underline{a}}{}^{[\overline{\cal C}}{}_{\overline{\cal E}} \; 
{\cal F}_{\underline{b}]}{}^{\overline{\cal D}]\overline{\cal E}}\right)\ ,\label{GenRim}
\end{eqnarray}
where $D_{\underline{a}}= \hat{e}^{m}{}_{\underline{a}} \partial_{m}$ and $\hat{e}$ is a local frame for the metric $\hat{g}$. 

There are also higher-derivative interactions, in implicit form, within the 3-form $\hat{H}=db-\hat\Omega_{CS}$, with the Lorentz Chern-Simons term $\hat\Omega_{CS}$
\begin{eqnarray}
(\hat\Omega_{CS})_{\underline{abc}}= 
\frac{\sqrt{2}}{g^2 X_R}{\cal F}_{[\underline{a}}{}^{\overline{\cal{CD}}}  
\left( 
\frac{3}{2} {\cal R}^{(+)}_{\underline{bc}]\overline{\cal CD}} 
+ 2 {\cal F}_{\underline{b}\overline{\cal CB}}
{\cal F}_{\underline{c}]\overline{\cal D}}{}^{\overline{\cal B}}
\right)\; .
\end{eqnarray}

A perturbative expansion in derivatives of (\ref{FullAction}) gives
\begin{eqnarray}
{\cal L}&=& \sqrt{-\hat{g}}\ e^{-2\hat{\phi}}  \left(\vphantom{\frac{e^{\frac12}}{2}}\right. \hat{R} - 4\ (\nabla \hat\phi)^2 + 4\  \Box\hat\phi -\frac{1}{12} \hat{H}_{mnp} \hat{H}^{mnp} 
\ + \ \frac{b}{2}
R^{(+)}_{\underline{abef}} 
R^{(+)}{}^{\underline{abef}} \cr
&&+
b^2 \left[ \hat{\cal D}^{(+)}{}^{[\underline{a}} R^{(+)}{}^{\underline{b}]\underline{ecd}}
\hat{\cal D}^{(+)}_{\underline{a}} R^{(+)}_{\underline{b}\underline{ecd}} 
+R^{(+)}{}^{\underline{abef}} \left( R^{(+)}_{\underline{aced}} 
R^{(+)}_{\underline{b}}{}^{\underline{cd}}{}_{\underline{f}}
-\frac12 R^{(+)}_{\underline{ae}}{}^{\underline{cd}} R^{(+)}_{\underline{bfcd}} \right)
\right] \left.\vphantom{\frac{e^{\frac12}}{2}}\right)\cr\cr
&& \;\;\;\;\;\;\;\;\;\;\;\;\;\;\;\;\;\;\;\;\;\;\;\;\;\;\;\;\;\;\;\;\;\;\;\;\;
 \;\;\;\;\;\;\;\;\;\;\;\;\;\;\;\;\;\;\;\;\;\;\;\;\;\;\;\;\;\;\;\;\;\;\;\;\;\;
 \;\;\;\;\;\;\;\;\;\;\;\;\;\;\;\;\;\;\;\;\;\;\;\;\;\;\;\;\;\;
+ {\cal O}(\alpha'{}^3)\; ,
\end{eqnarray}
where we have introduced
\begin{eqnarray}
\hat{\cal D}^{(+)}_{\underline{a}} 
R^{(+)}_{\underline{b}\underline{e}}{}^{\underline{cd}} 
&=&
D_{\underline{a}} {R}{}^{(+)}_{\underline{b}\underline{e}}{}^{\underline{cd}} 
+ \omega_{\underline{ab}}{}^{\underline{f}} 
 {R}{}^{(+)}_{\underline{fe}}{}^{\underline{cd}} 
 + \omega^{(+)}_{\underline{a}\underline{e}}{}^{\underline{f}} 
 {R}{}^{(+)}_{\underline{b}\underline{f}}{}^{\underline{cd}} 
 +  \omega^{(+)}_{\underline{a}}{}^{\underline{c}}{}_{\underline{f}} \ 
 {R}^{(+)}_{\underline{b}\underline{e}}{}^{\underline{fd}}
 +  \omega^{(+)}_{\underline{a}}{}^{\underline{d}}{}_{\underline{f}} \ 
 {R}^{(+)}_{\underline{b}\underline{e}}{}^{\underline{c}\underline{f}} 
 \ + \ {\cal O}(\alpha')
 \cr
 &=&
\nabla_{\underline{a}}R^{(+)}_{\underline{b}\underline{e}}{}^{\overline{cd}}
+\frac12 \hat{H}_{\underline{ae}}{}^{\underline{f}} R^{(+)}_{\underline{bf}}{}^{\underline{cd}} + \hat{H}_{\underline{af}}{}^{[\underline{c}} R^{(+)}_{\underline{be}}{}^{\underline{d}]\underline{f}}
 \ + \ {\cal O}(\alpha')\; .\label{DR+}
\end{eqnarray}

The remainder of this article is structured as follows. In Section~2, we review in some detail the original formulation of the generalized Bergshoeff-de Roo (gBdR) identification. In Section 3, we extend this identification to a Lorentz-covariant frame by employing a different parametrization of the generalized vielbein and by imposing gauge-fixing conditions that break the extended Lorentz group to the standard SUGRA Lorentz group from the outset. The generalized Riemann tensor and the corresponding action are then introduced.

In Section 4, we explicitly carry out the derivative expansion up to sixth order in derivatives. In Section 5, we perform some consistency checks of the resulting couplings. Finally, in Section 6, we present our conclusions and outline possible future directions. The work concludes with appendices, which provide a self-contained discussion of coset-representative parameterizations and collect auxiliary computational details.

\section{The generalized Bergshoeff-de Roo identification in the T-duality covariant framework}

The starting point in the gBdR identification is an extension of the heterotic formulation of DFT \cite{Hohm:2011nu} with duality group ${\cal G}=O(D+p,D+q)$, with $p$ and $q$ counting the number of negative and positive eigenvalues of the Killing metric of a certain gauge group $\cal K$.

The field content of the theory is the generalized dilaton $d$ and a generalized frame $\cal E_{\cal M}{}^{\cal A}$, the latter being a coset representative of ${\cal G}/{\cal H}$. Here ${\cal H}= \underline{\cal H}\times \overline{\cal H}$ is the extended Lorentz group, with $\underline{\cal H}=O(D-1,1)$ and $\overline{\cal H}=O(1+p,D+q-1)$. The generalized frame then preserves the invariant metric of ${\cal G}$, $\eta$.
\begin{eqnarray}
\eta_{\cal MN}= {\cal E}_{\cal M}{}^{\cal A}\  \eta_{\cal AB}\ 
{\cal E}_{\cal N}{}^{\cal B}  
\end{eqnarray}
with
\begin{eqnarray}
\eta_{\cal M N} = \left(\begin{matrix}\eta_{M N} & 0 \\ 0 & \kappa_{\tilde\mu \tilde\nu}\end{matrix}\right)\; ,\;\;\;\;\;\;\;\;\;\;\;
\eta_{\cal A B} = \left(\begin{matrix}\eta_{AB} & 0 \\ 0 & \kappa_{\overline{\alpha}\overline{\beta}}\end{matrix}\right)\; .\label{eta}
\end{eqnarray}
Here $\eta_{M N}$ and $\eta_{AB}$ are the invariant metric of $G=O(D,D)$, respectively
\begin{eqnarray} 
\eta_{M N} = \left(\begin{matrix} 0 & \delta^{m}{}_{n} \\ \delta_{m}{}^{n} & 0\end{matrix}\right)\; ,\;\;\;\;\;\;\;\;\;\;\;
\eta_{A B} = \left(\begin{matrix} - g_{\underline{ab}} & 0 \\ 0 & g_{\overline{ab}}\end{matrix}\right)\; ,
\end{eqnarray}
with $g_{\underline{ab}}$ and $g_{\overline{ab}}$ the flat Minkowski metric in $D$ dimensions and $\kappa$ being the Killing metric of the gauge algebra. 

The local symmetries are generalized diffeomorphisms and the extended Lorentz group
\begin{eqnarray}
\delta d &=& \xi^{\cal N} \partial_{N}d - \frac12 \partial_{\cal N} \xi^{\cal N} \ , \cr 
\delta {\cal E}_{\cal M}{}^{\cal A} &=& \xi^{\cal N}\partial_{\cal N}{\cal E}_{\cal M}{}^{\cal A} 
+  \left( \partial_{\cal M}\xi^{\cal N} - \partial^{\cal N}\xi_{\cal M} \right) {\cal E}_{\cal N}{}^{\cal A} 
+{\hat f}_{\cal M N}{}^{\cal P} \xi^{\cal N} {\cal E}_{\cal P}{}^{\cal A} 
+ {\cal E}_{\cal M}{}^{\cal B} \Gamma_{\cal B}{}^{\cal A} \;. \label{LocalTransformation}
\end{eqnarray}
The gauge symmetry must preserve $\eta$ and satisfy Jacobi identities. These conditions are expressed in terms of the following linear and quadratic constraints
\begin{eqnarray} 
 \hat{f}_{\cal M N P} = \hat{f}_{[{\cal M N P}]} \ , \ \ \ \ \hat{f}_{[{\cal M N}}{}^{\cal K} \hat{f}_{{\cal P}]{\cal K}}{}^{\cal L} = 0\ . 
 \end{eqnarray}
In addition, closure of the algebra requires the strong constraint
\begin{eqnarray} 
\eta^{\cal M N} \partial_{\cal M}\otimes \partial_{\cal N} =0\;,\;\;\;\;\;
\hat{f}_{\cal M N}{}^{\cal P} \partial_{\cal P}=0 \;.\label{SC}
\end{eqnarray}

The so-called generalized fluxes are given by
\begin{eqnarray}
{\cal F}_{\cal A}&=& 2 \;{\cal D}_{\cal A}d -\Omega_{\cal B A}{}^{\cal B}\, , \\
{\cal F}_{\cal A B C}&=& 
3\;\Omega_{[{\cal A B C}]} 
+ {\hat f}_{\cal M N P} {\cal E}^{\cal M}{}_{\cal A} {\cal E}^{\cal N}{}_{\cal B} {\cal E}^{\cal P}{}_{\cal C} \ , \label{ExtendedFluxes}
\end{eqnarray}
with ${\cal D}_{\cal A}= {\cal E}^{\cal M}{}_{\cal A} \partial_{\cal M}$ and $\Omega_{\cal A B C}$ being the generalized Weitzenb\"ock connection
\begin{eqnarray}
\Omega_{\cal A B C}= {\cal D}_{\cal A}{\cal E}^{\cal N}{}_{\cal B} {\cal E}^{\cal P}{}_{\cal C} \; \eta_{\cal N P} \ .
\end{eqnarray}
These fluxes are ${\cal G}$-invariant and transform under local transformations as
\begin{eqnarray}
\delta {\cal F}_{\cal A}&=&\xi^{\cal M}\partial_{\cal M}{\cal F}_{\cal A} + \Gamma^{\cal B}{}_{\cal A}\; {\cal F}_{\cal B} 
- {\cal D}_{\cal B}\Gamma^{\cal B}{}_{\cal A}\;,\\
\delta {\cal F}_{\cal A B C}&=&\xi^{\cal M}\partial_{\cal M}{\cal F}_{\cal A B C} \;
+ 3\;\left(\vphantom{\frac12}\Gamma^{D}{}_{[\cal A}\; {\cal F}_{\cal B C] D}- {\cal D}_{[\cal A} \Gamma_{\cal BC]}\right)\,. \label{TransfFluxes}
\end{eqnarray}
Under generalized diffeomorphisms they behave as scalars, while under Lorentz transformations they transform as connections. It is precisely these dual properties that elevate generalized fluxes to a central role in the flux formulation of DFT. In this formulation, $O(D,D+k)$ and generalized diffeomorphisms are manifest symmetries; Lorentz symmetry, however, is not manifest and instead serves to determine the coefficients of the action.

The coset representative of $O(D,D+k)/O(D-1,1)\times O(1,D+k-1)$ can be parametrized as
\begin{eqnarray}
{\cal E}_{{\cal M}}{}^{\cal A}=
\left(
\begin{matrix}
(\chi^{\frac12})_M{}^{N} E_{N}{}^{A}&  - {\cal A}_{M}{}^{\tilde\nu} e_{\nu}{}^{\overline \alpha}   \cr
{\cal A}^{M}{}_{\mu} E_{M}{}^{A} & 
(\Box^{\frac12})_{\tilde\mu}{}^{\tilde\nu} e_{\nu}{}^{\overline \alpha}
\end{matrix}
\right)\;,\label{ParDFT}
\end{eqnarray}
where $e_{\tilde\nu}{}^{\overline{\alpha}}$ is a local frame for the Killing metric $\kappa_{\tilde\mu\tilde\nu}$,  $E_{M}{}^{A}\in O(D,D)$ is
\begin{eqnarray}
E_{M}{}^{A}=
\frac{1}{\sqrt2}\left(
\begin{matrix}
\bar{e}^{m}{}^{\underline a} & \bar{e}^{m}{}^{\overline a}\cr
(\bar{b}_{mn} - \bar{g}_{mn})\bar{e}^{n \underline a} & 
(\bar{b}_{mn} + \bar{g}_{mn})\bar{e}^{n \overline a}\label{DFTparamet} \end{matrix}
\right)\;,
\end{eqnarray}
with
\begin{eqnarray}
\bar{g}^{mn}=\bar{e}^{m}{}_{\underline a} g^{\underline{ab}}\bar{e}^{n}{}_{\underline b}
=\bar{e}^{m}{}_{\overline a} g^{\overline{ab}}\bar{e}^{n}{}_{\overline b} ,    
\end{eqnarray}
and
\begin{eqnarray}
\chi_{M N} &=& \eta_{MN} - {\cal A}_{M}{}^{\tilde\nu} {\cal A}_{N\tilde\nu} \;, \cr
\Box_{\tilde\mu\tilde\nu} &=& \kappa_{\tilde\mu\tilde\nu} - {\cal A}_{M \tilde\mu} {\cal A}^{M}{}_{\tilde\nu} \;.
\end{eqnarray}
The tensors $\chi$ and $\Box$ satisfy a useful identity, which will be used repeatedly throughout this work:
\begin{eqnarray}
{\cal A}_{M}{}^{\mu} f(\Box)_{\mu}{}^{\nu} = f(\chi)_{M}{}^{N} {\cal A}_{N}{}^{\nu} .
\end{eqnarray}
This identity holds for any function $f$, though in practice we apply it to (fractional) powers of the argument.

The extended frame ${\cal E}\in{\cal G}/{\cal H}$ contains $D(D+k)$ degrees of freedom, which are parametrized in terms of the $D^2$ degrees of freedom of $E\in{O(D,D)/O(D-1,1)\times O(1,D-1)}$ and $Dk$ captured by ${\cal E}_{\tilde\mu}{}^{\underline{a}}$. The remaining components are non-physical and can be eliminated by the gauge fixing condition
\begin{equation} \label{gaugefixing}
{\cal E}^{\tilde\nu}{}_{\overline a}=E^{M}{}_{\overline a} \, {\cal A}_{M}{}^{\tilde\nu}=0 \ , \ \ \  e_{\tilde\mu}{}^{\overline\alpha}={\rm constant} \,. 
\end{equation}
This requirement implies $\delta {\cal E}^{\tilde\nu}{}_{\overline a}=0=\delta e_{\tilde\mu}{}^{\overline\alpha}$ which fixes $\Gamma_{\overline{\alpha\beta}}$ and $\Gamma_{\overline{a\beta}}$ 
\begin{eqnarray}
\Gamma_{\overline {\alpha a}} &=& e^{\tilde\mu}{}_{\overline \alpha} \, (\Box^{- \frac 1 2}){}_{\tilde\mu}{}^{\tilde\nu}\, \partial_M \xi_{\tilde\nu} \, E^M{}_{\bar a}\ , \label{GaugedFixedParametersMono}\\
\Gamma_{\overline {\alpha \beta}} &=&  e^{\tilde\mu}{}_{[\overline \alpha} \, e^{\tilde\nu}{}_{\overline \beta]} \, (\Box^{-\frac 1 2}){}_{\tilde\mu}{}^{\tilde\rho} \left(\delta (\Box^{\frac 1 2}){}_{\tilde\rho \tilde\nu} - {\cal A}^{M}{}_{\tilde\nu}  \, \partial_{M} \xi_{\tilde\rho}- g\, f_{ \tilde\rho \tilde\sigma}{}^{\tilde\tau} \, \xi^{\tilde\sigma} \,(\Box^{\frac 1 2}){}_{\tilde\tau \tilde\nu} \right)
\ , \nonumber
\end{eqnarray}
and so breaks the extended Lorentz group ${\cal H}=O(D-1,1) \times O(1,D+k-1)$ into the double one $H=O(D-1,1) \times O(1,D-1)$. This is an important step of the construction because, as we commented in the introduction, the gauge group is auxiliary and non-physical, so we must end up at the end with the degrees of freedom and gauge groups of the ungauged DFT.  The final step is to fix the $Dk$ extra degrees of freedom and it is done by identifying the gauge vectors in ${\cal E}^{\tilde\nu}{}_{\underline a}$ with ${\cal F}_{\underline{a}\overline{\cal CD}}$.

In order to proceed, let us notice that
\begin{eqnarray}
\delta {\cal E}_{\tilde\mu \underline a } = \xi^{N}\partial_{N}{\cal E}_{\tilde\mu }{}^{\underline{a}}  - {\cal D}_{\underline a} \xi_{\tilde\mu} + g f_{\tilde\mu \tilde\nu}{}^{\tilde\rho} \xi^{\tilde\nu} {\cal E}_{\tilde\rho \underline a} + {\cal E}_{\tilde\mu \underline d} \Gamma^{\underline d}{}_{\underline {a}} \ . \label{TransformationA}
\end{eqnarray}
Here we considered
\begin{eqnarray}
\hat f_{\cal M N}{}^{\cal P}= \left\{  \begin{array}{cc}
  g \ f_{\tilde\mu\tilde\nu}{}^{\tilde\rho},& \ {\rm for} \ {\cal M, N, P}=\tilde\mu,\tilde\nu,\tilde\rho   \\
  0,&   \ {\rm otherwise}    
\end{array}
\right.  
\end{eqnarray}
with $f_{\tilde\mu\tilde\nu}{}^{\tilde\rho}$ being the structure constants of the gauge group ${\cal K}$ and $g$ the gauge coupling. Then the section condition (\ref{SC}) turns into
\begin{eqnarray}
\partial_{\tilde\mu}=0\; , \;\;\;\;\;\;\;\;\;\;\; \eta^{MN} \partial_{M}\otimes \partial_{N}=0   \;.
\end{eqnarray}

The hint is in introducing (\ref{IndexRelation}) which leads to (\ref{TransformationAH}) suggesting that the auxiliary gauge group is fixed by
\begin{eqnarray}
{\cal K}=\overline{\cal H}  \;.  
\end{eqnarray}

There is a subtlety in this identification as ${\rm dim}({\cal K})=k\neq (D+k)(D+k-1)/2={\rm dim}(\bar{\cal H})$ for any finite $k$, so an isomorphism is only possible for infinite $k$. Despite this observation, the computations can be performed without specifying the value of $k$, so one can proceed and take the limit $k\to \infty$ at the end. 

The generators of the gauge algebra play a central role in the identification. They satisfy the following relations 
\begin{eqnarray}
(t^{\tilde\mu})_{\overline{\cal AB}} (t_{\tilde\nu})^{\overline{\cal AB}} = X_R \ \delta^{\tilde{\mu}}_{\tilde{\nu}} \;, \;\;\;\;\;\;\;\;\;\;\;\;\;\;
(t^{\tilde\mu})_{\overline{\cal AB}} (t_{\tilde\mu})^{\overline{\cal CD}} = X_R \ \delta^{\overline{\cal AB}}_{\overline{\cal CD}}\;, \;\label{normalization}
\end{eqnarray}
$X_R$ being the Dynkin index of the representation and $\overline{\cal A}=\{\overline{a},\overline{\alpha}\}$. Gauge indices are (raised) lowered with the (inverse) Killing metric. For instance, $t^{\tilde\mu}:=\kappa^{\tilde\mu\tilde\nu} t_{\tilde\nu}$ and $f_{\tilde\mu\tilde\nu\tilde\rho}:=\kappa_{\tilde\rho\tilde\sigma} f_{\tilde\mu\tilde\nu}{}^{\tilde\sigma}$.

The identification ${\cal F}_{\underline a \overline {\cal B C}} = -g  \, {\cal E}_{\tilde\mu \underline a} \, (t^{\tilde\mu})_{\overline {\cal B C}}$ is dubbed Generalized Bergshoeff-de Roo identification. It triggers an iterative process in the action; indeed, the generalized fluxes contain, after the parameterization (\ref{ParDFT}), derivatives of the generalized vielbein $E_{M}{}^{A}$, leading to fluxes of the ungauged DFT, as well as derivatives of ${\cal E}_{\tilde\mu \underline a}$, which can be written, once again, in terms of derivatives of ${\cal F}_{\underline a \overline {\cal B C}}$ after the identification. Repeating this, one always obtains an expansion containing derivatives of the ungauged generalized fluxes plus higher-order derivatives of ${\cal F}_{\underline a \overline {\cal B C}}$. 

Similarly, the identification $\xi^{\tilde\mu}\to \Gamma^{\overline{BC}}$ renders an iterative process leading to Green-Schwarz transformations of the generalized frame. 

For instance, this mechanism leads to (see \cite{Baron:2018lve} for further details)
\begin{eqnarray}
\delta E_{M}{}^{\overline a}&=& \ \widehat {\cal L}_\xi E_M{}^{\overline a} + E_M{}^{\overline b}\, \Lambda_{\overline b}{}^{\overline a} - \frac b 2  \, E_M{}^{\underline d} F_{\underline d \overline {b c}} \, D^{\overline a}  \Lambda^{\overline {bc}}  + {\cal O}(b^2)\; ,
\cr
\delta E_M{}^{\underline a} &=& \ \widehat {\cal L}_\xi E_M{}^{\underline a} + E_M{}^{\underline b}\, \Lambda_{\underline b}{}^{\underline a} + \frac b 2 \partial_{\overline M} \Lambda^{\overline {b c}} \, F^{\underline a}{}_{\overline {b c}} 
 + {\cal O}(b^2)\; ,
\end{eqnarray}
where the parameter $b\sim g^{-2} \sim \alpha'$. Here the redefined Lorentz parameters are
\begin{eqnarray}
\Lambda_{\underline {a b}} &=& \Gamma_{\underline {a b}} - E^M{}_{[\underline a} E^N{}_{\underline b]} (\chi^{-\frac 1 2}){}_M{}^P \left(\delta (\chi^{\frac 1 2})_{P N} - \partial_P \xi^\alpha {\cal A}_{N \alpha}\right)\ , \label{LorentzParamterRedefinition}\\
\Lambda_{\overline {a  b}} &=& \Gamma_{\overline {a b}}  \ . \nonumber
\end{eqnarray}
This is the first order of the generalized Green-Schwarz transformation \cite{Marques:2015vua}. Using the parametrization (\ref{DFTparamet}) one obtains the well-known Green-Schwarz transformation of the Kalb-Ramond field $\bar{b}_{mn}$. In addition, it leads to a correction in the Lorentz transformation of $\bar{e}_{m}{}^{\underline{a}}$ and $\bar{e}_{m}{}^{\overline{a}}$, so that $\bar{g}_{mn}$ is no longer Lorentz invariant. This is a manifestation that field redefinitions are needed at higher orders in derivatives to relate duality covariant multiplets with Lorentz covariant ones.

The action derived from the direct application of the gBdR identification consists of 8 terms at $\mathcal{O}(1)$ in the frame formulation of DFT, 16 couplings at $\mathcal{O}(\alpha')$, and roughly 300 terms at $\mathcal{O}(\alpha'^2)$. Consequently, the field redefinition required to cast the action into a standard supergravity (SUGRA) scheme quickly becomes cumbersome. For instance, while $\bar{e}$ is identified with the physical (Lorentz-covariant) vielbein at leading order, higher-order field redefinitions introduce 1 additional term at $\mathcal{O}(\alpha')$ and approximately 30 terms at $\mathcal{O}(\alpha'^2)$.

Despite this complexity, further simplifications lead to compact expressions for the action at higher orders, and partial checks were early performed in \cite{Hronek:2021nqk}. Moreover, the authors of \cite{Hronek:2022dyr} constructed the $\mathcal{O}(\alpha'^2)$ action in DFT without invoking the gBdR approach. Instead, they started from the $\mathcal{O}(\alpha')$ action, incorporated the known correction to the double Lorentz transformations, and required invariance at sixth order in derivatives, thereby explicitly verifying the agreement between both methods.

In this work, we revisit the explicit computation of these couplings at ${\cal O}(\alpha'{}^2)$ by introducing a modification to the original approach that bypasses the need for field redefinitions. 

An alternative approach, based on the twisted Pol\'a\v{c}ek-Siegel construction, was recently presented in~\cite{Gitsis:2024gfb}. Building on earlier ideas~\cite{Polacek:2013nla,Butter:2022iza,Gitsis:2025clo}, this method employs an extended space, the \textit{mega-space}, and offers a geometric reinterpretation of the gBdR identification. A subsequent refinement in~\cite{Gitsis:2026ufz}, achieved through a suitable choice of algebra-generator basis, greatly simplifies the construction and makes its computational implementation much more accessible. This decomposition naturally yields a Lorentz-covariant frame, and it would be worthwhile to investigate its connection to the approach developed here.

\section{The generalized Bergshoeff-de Roo identification in the Lorentz covariant framework}

The coset representative of $O(D,D+k)/O(D-1,1)\times O(1,D+k-1)$ can be parametrized alternatively as follows 
\begin{eqnarray}
{\cal E}_{{\cal M}}{}^{\cal A}=
\left(
\begin{matrix}
\frac{1}{\sqrt 2} \hat{e}^{m \underline{a}} & 
\frac{1}{\sqrt 2} \hat{e}^{m \overline{a}} &
\frac{1}{\sqrt 2} \hat{e}^{m \overline{\alpha}} \cr
-\frac{1}{\sqrt 2} \left(\hat{c}_{pm}+ \hat{g}_{pm}\right)\hat{e}^{p \underline{a}} & 
- \frac{1}{\sqrt 2} \left(\hat{c}_{pm} - \hat{g}_{pm}\right)\hat{e}^{p \overline{a}} + \hat{A}_{m}{}^{\overline{a}}&
- \frac{1}{\sqrt 2} \left(\hat{c}_{pm} - \hat{g}_{pm}\right)\hat{e}^{p \overline{\alpha}} + \hat{A}_{m}{}^{\overline{\alpha}}\cr
- \frac{1}{\sqrt 2}\hat{A}_{p\tilde\mu} \hat{e}^{p \underline{a}} & 
- \frac{1}{\sqrt 2} \hat{A}_{p\tilde\mu}\hat{e}^{p\overline{a}} + \hat{e}_{\tilde\mu}{}^{\overline{a}}&
- \frac{1}{\sqrt 2} \hat A_{p\tilde\mu} \hat{e}^{p\overline{\alpha}}+ \hat{e}_{\tilde\mu}{}^{\overline{\alpha}}
\end{matrix}
\right)\; ,\label{ParHet}
\end{eqnarray}
where 
\begin{eqnarray}
\hat A_{m \tilde\mu}&\equiv& \hat A_{m}{}^{\overline{\cal A}} \hat e_{\tilde\mu {\overline{\cal A}}} = \hat A_{m}{}^{\overline{a}} \hat e_{\tilde\mu {\overline{a}}} + \hat A_{m}{}^{\overline{\alpha}} \hat e_{\tilde\mu {\overline{\alpha}}}\; ,\cr
\hat c_{mn} &\equiv& \hat{b}_{mn} +\frac12 \hat A_{m \tilde\mu} \hat A_{n}{}^{\tilde\mu}\; ,\cr
\hat g^{mn} &=& \hat e^{m}{}_{\underline a}\ g^{\underline{ab}} \ \hat e^{n}{}_{\underline b} = \hat e^{m}{}_{\overline a}\ g^{\overline{ab}} \ \hat e^{n}{}_{\overline b} + 
\hat e^{m}{}_{\overline \alpha}\ \kappa^{\overline{\alpha\beta}} \ \hat e^{n}{}_{\overline \beta}\; ,  \cr
\kappa_{\tilde\mu\tilde\nu} &\equiv& \hat e_{\tilde\mu}{}^{\overline{ \cal A}} e_{\tilde\nu \overline{\cal A}} 
= \hat e_{\tilde\mu}{}^{\overline a} \hat e_{\tilde\nu \overline a} + \hat e_{\tilde\mu}{}^{\overline \alpha} \hat e_{\tilde\nu \overline \alpha}\; . \label{CF}
\end{eqnarray}

We use the convention that $O(D,D)$ indices are lowered and raised with $\hat \eta_{\cal AB},\ \hat \eta_{\cal MN}$ and $\hat \eta^{\cal AB},\ \hat \eta^{\cal MN}$, while $GL(D)$ and $GL(k)$ indices are lowered with $g_{\underline{ab}},\ g_{\overline{ab}},\ g_{mn}$ and $\kappa_{\overline{\alpha\beta}}, \kappa_{\mu\nu}$, respectively and raised with their inverses.\footnote{The reader should carefully work expressions containing $O(D,D)$ tensors and $GL(D)$ tensors.  For instance, ${\cal E}^{m\underline{a}}=\frac{1}{\sqrt{2}} \hat{e}^{m\underline{a}}$, while ${\cal E}^{m}{}_{\underline{a}}=\eta_{\underline{ab}}{\cal E}^{m}{}^{\underline{b}}=-g_{\underline{ab}}{\cal E}^{m}{}^{\underline{b}}=- \frac{1}{\sqrt{2}} \hat{e}^{m}{}_{\underline{a}}$.} Here $g_{\underline{ab}}= \delta_{\underline a}^{c}\  \delta_{\underline b}^{d} \ g_{cd}$ and $g_{\overline{ab}}= \delta_{\overline a}^{c} \ \delta_{\overline b}^{d} \ g_{cd}$, with $g_{cd}$ being the Minkowski metric. For instance, while $\hat e^{m}{}_{\underline a}$ is the inverse transpose of $\hat e_{m}{}^{\underline a}$, $\hat e^{m}{}_{\overline a} = \hat g^{mn} \ g_{\overline{ab}} \hat e_{n}{}^{\overline b}$ is not the inverse transpose of $\hat e_{m}{}^{\overline a}$.
 
The first relation in (\ref{CF}) can be inverted 
\begin{eqnarray}
\hat{A}_{m\overline{a}} =\hat{A}_{m}{}^{\tilde\mu} \hat{e}_{\tilde\mu \overline{a}}  \;\;,\;\;\;\;\;\;\;\;\;\;\;\;
\hat{A}_{m\overline{\alpha}} =\hat{A}_{m}{}^{\tilde\mu} \hat{e}_{\tilde\mu \overline{\alpha}}   \; .
\end{eqnarray}

The gauge field is constrained so as to satisfy 
\begin{eqnarray}
 0 = \hat A_{m \overline{\cal A}} \ \hat e_{n}{}^{ \overline{\cal A}}
 = \hat A_{m \overline{a}} \ \hat e_{n}{}^{ \overline{a}} +
 \hat A_{m \overline{\alpha}} \ \hat e_{n}{}^{ \overline{\alpha}} \; , \label{Id1}
\end{eqnarray}
and $\hat e_{\tilde\mu}{}^{\overline a},\ \hat e_{\tilde\mu}{}^{\overline \alpha}, \ \hat e_{m}{}^{\overline a},\ \hat e_{m}{}^{\overline \alpha}$ further satisfy
\begin{eqnarray}\cr
 0 &=& \hat e_{m \overline{\cal A}} \ \hat e_{\tilde\nu}{}^{ \overline{\cal A}}
 = \hat e_{m \overline{a}} \ \hat e_{\tilde\nu}{}^{ \overline{a}} +
 \hat e_{m \overline{\alpha}} \ \hat e_{\tilde\nu}{}^{ \overline{\alpha}}  \label{Id2}  
\end{eqnarray}
which together with (\ref{CF}) imply
\begin{eqnarray}
\hat e_{\widetilde M}{}^{\overline{\cal A}}\ \hat e^{\widetilde N}{}_{\overline{\cal A}} = \delta_{\widetilde M}{}^{\widetilde N}\;, \label{Id3} \\
\hat e_{\widetilde M}{}^{\overline{\cal A}}\ \hat e^{\widetilde M}{}_{\overline{\cal B}} = \delta^{\overline{\cal A}}{}_{\overline{\cal B}}\; , \label{Id4}
\end{eqnarray}
with $\widetilde M=\{m,\mu\}$. 
\bigskip

The field $\hat{e}^{m\underline a}$ in (\ref{ParHet}) is related to $E_{N}{}^{A}$ in (\ref{ParDFT}) through
\begin{eqnarray}
\hat{e}^{m\underline a} = \sqrt{2}\ (\chi^{\frac12})^{m}{}_{N} E^{N}{}^{\underline a}\; .   
\end{eqnarray}

Consequently, we can relate $\hat{g}^{mn}$ and $\overline{g}^{mn}$ 
\begin{eqnarray}
\hat{g}^{mn} &=& \hat{e}^{m\underline a} \hat{e}^{n}{}^{\underline b} g_{\underline{ab}} = - \hat{e}^{m\underline a} \hat{e}^{n}{}^{\underline b} \eta_{\underline{ab}}\cr
&=& -2 \  \eta_{\underline{ab}} \ (\chi^{\frac12})^{m}{}_{M} E^{M}{}^{\underline a} (\chi^{\frac12})^{n}{}_{N} E^{N}{}^{\underline b} \cr
&=& -2 \  \eta_{AB} \ (\chi^{\frac12})^{m}{}_{M} E^{M}{}^{A} (\chi^{\frac12})^{n}{}_{N} E^{N}{}^{B} 
+ 2 \  \eta_{\overline{ab}} \ (\chi^{\frac12})^{m}{}_{M} E^{M}{}^{\overline a} (\chi^{\frac12})^{n}{}_{N} E^{N}{}^{\overline b} \cr
&=& -\ 2\ (\chi^{\frac12})^{m}{}_{M} E^{M}{}^{A} (\chi^{\frac12})^{n}{}_{N} E^{N}{}_{A}
+\ 2\ (\chi^{\frac12})^{m}{}_{M} E^{M}{}^{\overline a} (\chi^{\frac12})^{n}{}_{N} E^{N}{}_{\overline a} \cr
&=& -2 (\chi^{\frac12})^{m}{}_{M} \eta^{MN} (\chi^{\frac12})^{n}{}_{N} + \overline{e}^{m \overline a} \overline{e}^{n}{}_{\overline a} \cr
&=&\overline{g}^{mn} - 2\ \chi^{mn}\cr
&=& \overline{g}^{mn} + 2\ {\cal A}^{m}{}_{\tilde\mu}{\cal A}^{n}{}^{\tilde\mu}\;.\label{hgTobg1}
\end{eqnarray}

Alternatively, if we impose the gauge condition ${\cal E}_{\tilde\mu}{}^{\overline a}=0$, the previous expression can be put in the form:
\begin{eqnarray}
\hat{g}^{mn}= \bar{g}^{mn} 
+ {\cal E}_{\tilde\mu}{}^{\underline a}    
{\cal E}^{\tilde\mu}{}^{\underline b} \overline{e}^{m}{}_{\underline a} \overline{e}^{n}{}_{\underline b} \;.\label{hgTobg2} 
\end{eqnarray}

Interestingly, this expression exactly matches the relation between $\bar{g}^{mn}$ and $g^{mn}$, the physical (Lorentz invariant) metric obtained in \cite{Baron:2021yqm} (see eq. (4.15)), if we identify $\hat{g}^{mn}=g^{mn}$. 

In the rest of this section, we will independently demonstrate that $\hat g^{mn}$ is Lorentz invariant. Furthermore, we will show that $\hat b_{mn}$ is not a Lorentz singlet—instead, it transforms precisely as dictated by the Green-Schwarz mechanism. Additionally, we will prove that its curvature, $\hat H_{mnp}$, is all-order Lorentz invariant.

\subsection{Lorentz singlets}

The local transformation of the fields in (\ref{ParHet}), when the strong constraint is solved with the standard supergravity section $\partial_{{\cal M}}=(\partial^{\widetilde{m}},\partial_m,\partial_{\tilde\mu})=(0,\partial_{m},0)$ along with $\hat f_{\cal MN}{}^{\cal P}=g \ f_{\tilde\mu\tilde\nu}{}^{\tilde\rho}$ (and zero for all other index choices),\footnote{Here $g$ is the gauge coupling, not to be confused with the metric field $\hat{g}_{mn}$. $[g]={\rm length}^{-1}$, so $f_{\tilde\mu\tilde\nu\tilde\rho}$ are the standard dimensionless structure constants of the gauge algebra, while $\hat{f}$ has dimensions of flux.} readily follows after plugging (\ref{ParHet}) into (\ref{LocalTransformation}). For instance, $\delta {\cal E}^{m\underline{a}}$ implies
\begin{eqnarray}
 \delta \hat{e}^{m\underline{a}} = {\cal L}_{\xi^m} \hat{e}^{m \underline a} 
 +  \hat{e}^{m\underline b}\ \Gamma_{\underline b}{}^{\underline a}\;, \label{deltae}
 \end{eqnarray}
where 
\begin{eqnarray}
{\cal L}_{\xi^m} \hat{e}^{m \underline a}= 
\xi^{n} \partial_{n} \hat{e}^{m \underline a}
- \partial_{n} \xi^{m} \hat{e}^{n \underline a}  \;,  
\end{eqnarray}
is the Lie derivative of a vector. 

Considering instead $\delta {\cal E}_{\tilde\mu}{}^{\underline{a}}$ leads to 
\begin{eqnarray}
\delta \hat{A}_{m\tilde\mu} = {\cal L}_{\xi^n} \hat{A}_{m\tilde\mu} + \delta_{\xi^{\tilde\nu}} \hat{A}_{m\tilde\mu}  \; ,  \label{deltaA}
\end{eqnarray}
where ${\cal L}_{\xi^n} \hat{A}_{m\tilde\mu}= \xi^{n}\partial_{n} \hat{A}_{m\tilde\mu} + \partial_{m}\xi^{n} \hat{A}_{n\tilde\mu}$ is the Lie derivative of a 1-form and $\delta_{\xi^{\tilde\nu}}\hat{A}_{m\tilde\mu}=\partial_{m}\xi_{\tilde\mu} +\hat{f}_{\tilde\mu\tilde\nu}{}^{\tilde\rho} \xi^{\tilde\nu}\hat{A}_{m\tilde\rho}$ is the non-Abelian gauge transformation of the vector field.

The curvature of the $\hat{A}$ fields 
\begin{eqnarray}
\hat{F}_{mn\tilde\rho}= 2\ \partial_{[m}\hat{A}_{n]\tilde\rho} - \hat{f}_{\tilde\mu \tilde\nu \tilde\rho} \ \hat{A}_{m}{}^{\tilde\mu}\ \hat{A}_{n}^{\tilde\nu}    \label{Fmn}
\end{eqnarray}
is Lorentz invariant and it transforms covariantly under diffeomorphisms and gauge transformations:
\begin{eqnarray}
\delta \hat{F}_{mn\tilde\rho} ={\cal L}_{\xi^{p}} \hat{F}_{mn\tilde\rho} +
[\xi,\hat{F}_{mn}]_{\tilde\rho}\label{deltaF}
\end{eqnarray}
where ${\cal L}_{\xi^{p}} \hat{F}_{mn\tilde\rho}$ is the Lie derivative of a 2-form and 
$[\xi,\hat{F}_{mn}]_{\tilde\rho}= \hat{f}_{\tilde\mu \tilde\nu\tilde\rho} \xi^{\tilde\nu} \hat{F}_{mn}{}^{\tilde\mu}$.

Finally, $\delta{\cal E}_{m}{}^{\underline a}$ leads to
\begin{eqnarray}
\delta {\hat b_{mn}}=
{\cal L}_{\xi^{p}} \hat{b}_{mn} + \delta_{\xi_{p}}\hat{b}_{mn} + \delta_{\xi^{\tilde\nu}} \hat{b}_{mn}\;,\label{deltab}
\end{eqnarray}
where 
\begin{eqnarray}
{\cal L}_{\xi^{p}} \hat{b}_{mn} = 
\xi^{n}\partial_{n}\hat{b}_{mn} 
+ 2\ \partial_{[m}\xi^{p} \ \hat{b}_{n]p} 
\end{eqnarray}
is the Lie derivative of a 2-form, 
\begin{eqnarray}
\delta_{\xi_{p}}\hat{b}_{mn}= \ 2 \ \partial_{[m}\xi_{n]}
\end{eqnarray}
is the usual gauge transformation of the 2-form and
\begin{eqnarray}
\delta_{\xi^{\tilde\nu}} \hat{b}_{mn} = - \partial_{[m} \xi^{\tilde\nu} \hat{A}_{n]\tilde\nu} \label{deltabA}     
\end{eqnarray}
is the gauge transformation of the 2-form associated with the gauge field $\hat{A}_{m\tilde\nu}$. 

The curvature of the 2-form:
\begin{eqnarray}
\hat{H}_{mnp} = 3 \partial_{[m}\hat{b}_{np]} - \hat\Omega_{CS}(A)\;,\label{HSugra}
\end{eqnarray}
with the Chern-Simons term
\begin{eqnarray}
 \hat\Omega_{CS}(A)   = 3 \kappa_{\tilde\mu\tilde\nu} \hat{A}_{[m}{}^{\tilde\mu}\ \partial_{n}\hat{A}_{p]}{}^{\tilde\nu} \ 
- \ \hat{f}_{\tilde\mu\tilde\nu\tilde\rho} \ \hat{A}_{[m}{}^{\tilde\mu}\ \hat{A}_{n}{}^{\tilde\nu}\ \hat{A}_{p]}{}^{\tilde\rho}\label{OCS}
\end{eqnarray}
is Lorentz and gauge invariant
\begin{eqnarray}
\delta\hat{H}_{mnp}={\cal L}_{\xi^{q}}\hat{H}_{mnp}    
\end{eqnarray}
is the Lie derivative of a 3-form.
\bigskip

At this stage, $\hat{b}_{mn}$ appears to be Lorentz invariant. However, after implementing the gBdR identification (where we equate the gauge group with the Lorentz group) this identification induces a Green-Schwarz transformation on $\hat{b}$. Indeed, (\ref{deltabA}) can be reformulated as   
\begin{eqnarray}
\delta_{\xi^{\nu}}\hat{b}_{mn} = \sqrt{2} \partial_{[m}\xi^{\tilde\nu}\hat{e}_{n] \underline{a}} {\cal E}_{\tilde\nu}{}^{\underline a}  \;,  
\end{eqnarray}
which can be written, after the gBdR identification ($\xi^{\tilde\nu}\to\Gamma^{\overline{CD}},\ {\cal E}_{\tilde\nu}{}^{\underline{a}}\to {\cal F}^{\underline a}{}_{\overline{CD}}$) as
\begin{eqnarray}
\delta_{\Gamma}\hat{b}_{mn} = \frac{\sqrt{2}}{g^2\ X_R } \partial_{[m}\Gamma^{\overline{CD}}\hat{e}_{n] \underline{a}} {\cal F}^{\underline{a}}{}_{\overline{CD}}  \;,  
\end{eqnarray}
which leads after an iterative process to the first-order Green-Schwarz transformations (see \cite{Baron:2018lve} for details)
\begin{eqnarray}
\delta_{\Lambda}\hat{b}_{mn} = \frac{b}{2}\partial_{[m}\Lambda^{\overline{cd}}\omega^{(+)}_{n]\overline{cd}}\; + {\cal O}(\alpha'{}^2).  
\end{eqnarray}

The crucial observation here is that Lorentz transformations of the fields after gBdR identification enter either through the Lorentz transformation of Extended DFT $\delta_{\Gamma}$ or through the gauge transformations $\delta_{\xi^{\tilde\nu}}$. Both $\hat{g}_{mn}$ and $\hat{H}_{mnp}$ are invariant under both Lorentz and gauge transformations, which means they remain Lorentz-invariant after gBdR identification. Therefore, they correspond to the physical (Lorentz-singlet) degrees of freedom in the SUGRA scheme.

So far we have focused on the generalized vielbein while ignoring the dilaton. From the discussions in \cite{Baron:2018lve}, \cite{Baron:2020xel} and \cite{Baron:2021yqm}, we know that the generalized dilatons in Extended DFT, ungauged DFT, and the SUGRA scheme all agree. This means the dilaton in the SUGRA scheme is precisely the dilaton of Extended DFT, $\hat\phi$.

\subsection{Gauge fixing}
The goal of this article is to implement now the gBdR identification in such a way that the output is in the SUGRA scheme. From the previous discussion, the parametrization of the generalized frame is necessarily the one in (\ref{ParHet}). In order to do that, on top of this choice, we must also fix certain redundancies. This means that 
\begin{itemize}
    \item We will assume the SUGRA section to hold: 
\begin{eqnarray}
\partial_{{\cal M}}=(\partial^{\widetilde{m}},\partial_m,\partial_{\mu})=(0,\partial_{m},0) \;,\;\;\;\;\;\;\;\;\; {\hat f_{\cal MN}{}^{\cal P}}\to g f_{\tilde \mu \tilde\nu}{}^{\tilde\rho}   \label{SUGRAsection}    
\end{eqnarray} 
    \item We will break the Extended Lorentz group into the standard one: 
\begin{eqnarray}
O(D-1,1)\times O(p+1,D+q-1)\ \to \ O(1,D-1)    
\end{eqnarray}     
\end{itemize}

The latter is achieved by imposing 3 different gauge fixings. The first 2 are those of the gBdR identification discussed in the previous section, i.e. \\
\\
$1)\;\;\;\; {\cal E}_{\tilde\mu}{}^{\overline{a}}=0$, which now implies $e_{\tilde\mu}{}^{\overline a} = \frac{1}{\sqrt{2}} A_{p\tilde\mu} \hat{e}^{p\overline{a}}$.\\
 \\
$2)\;\;\;\; \overline{e}_{\tilde\mu}{}^{\alpha}={\rm constant}$, which implies a constant $\kappa_{\tilde\mu\tilde\nu}$.

These conditions fix the components $\Gamma^{\overline a \overline \alpha}$ and $\Gamma^{\overline{\alpha\beta}}$, leaving only $\Gamma^{\overline{ab}}$ and $\Gamma^{\underline{ab}}$ as free parameters. So, producing
\begin{eqnarray}
O(D-1,1)\times O(p+1,D+q-1)\ \to \ O(D-1,1)\times O(1,D-1)      
\end{eqnarray}

The condition ${\cal E}_{\tilde\mu}{}^{\overline{a}}=0$ puts some of the components of (\ref{ParHet}) in a more symmetric way, significantly simplifying the subsequent computation of the generalized fluxes. This condition requires:
\begin{eqnarray}
0&=&{\cal E}^{\cal N}{}_{\overline a} \ {\cal E}_{\cal N}{}_{\underline b}= {\cal E}^{m}{}_{\overline a} \ {\cal E}_{m}{}_{\underline b} +
{\cal E}_{m}{}_{\overline a} \ {\cal E}^{m}{}_{\underline b} \; .
\end{eqnarray}
Substituting the parametrization from (\ref{ParHet}) into the right-hand side leads to
\begin{eqnarray}
A_{m \overline{a}}= \frac{1}{\sqrt{2}} \ \hat{e}^{p}{}_{\overline a} \ A_{m\tilde\mu}  A_{p}{}^{\tilde\mu}    \; ,
\end{eqnarray}
which in turn yields
\begin{eqnarray}
{\cal E}_{m}{}^{\overline a}&=&  - \frac{1}{\sqrt 2} \left(\hat{c}_{pm} - \hat{g}_{pm}\right)\hat{e}^{p \overline{a}} + \hat{A}_{m}{}^{\overline{a}} \cr 
&=&  \frac{1}{\sqrt 2} \left(\hat{c}_{mp} + \hat{g}_{mp}\right)\hat{e}^{p \overline{a}}\; .
\end{eqnarray}

A third gauge-fixing condition is (see Section 4 of \cite{Baron:2020xel})\\
\\
$3) \;\;\;\;  \bar{e}^{m \overline a}= \delta^{\overline a}{}_{\underline b} \ \bar{e}^{m \underline b}\;.$ This requires $\Lambda_{\underline{ab}}\ =- \  \delta_{\underline{a}}^{\overline{c}} \ \delta_{\underline{b}}^{\overline{d}} \ \Lambda_{\overline{cd}}\ + {\cal O}(\alpha'{}^2)$.
\medskip

Here $\Lambda$ refers to the redefined parameters (\ref{LorentzParamterRedefinition}). This forces an identification between $O(D-1,1)$ and $O(1,D-1)$. Hence, producing 
\begin{eqnarray}
O(D-1,1)\times O(1,D-1) \ \to \ O(1,D-1)      
\end{eqnarray}

From earlier discussions, we identify $\hat{e}^{m \underline{a}}$ as the vielbein of the SUGRA scheme, while $\hat{e}^{m \overline{a}}$ is equal to $\bar{e}^{m \overline{a}}$ (one of the two vielbeins parameterizing the generalized vielbeins of DFT). Therefore, the last gauge condition translates into:
\begin{eqnarray}
\hat{e}^{m\underline a} &=& \delta^{\underline a}{}_{b} \ e^{mb}, \cr
\hat{e}^{m\overline a} &=& \delta^{\overline a}{}_{\underline c} \ (\chi^{-\frac12})^{\underline c}{}_{\underline d} \ \delta^{\underline d}{}_{b} \ e^{mb},\label{vielbeins}
\end{eqnarray}
where $e^{mb}$ is the standard vielbein of the supergravity scheme and we have introduced\footnote{Notice that $\chi^{\underline{a}}{}_{\underline{b}}$ is the only non trivial component, after the choice ${\cal E}_{\tilde\mu}{}^{\overline{a}}=0$. Indeed $\chi^{\overline{a}}{}_{\overline{b}}\to\delta^{\overline{a}}{}_{\overline{b}}$ and $\chi^{\overline{a}}{}_{\underline{b}},\chi^{\underline{a}}{}_{\overline{b}}\to 0$.} 
\begin{eqnarray}
\chi^{A}{}_{B}:=E_{M}{}^{A} E^{N}{}_{B} \chi^{M}{}_{N} \;.   
\end{eqnarray}
Regarding the 2-form, it is readily verified that
\begin{eqnarray}
\bar{b}_{mn}=\hat{b}_{mn}    \;.
\end{eqnarray}

From the discussion above and the relation
\begin{eqnarray}
\hat{c}_{mn}+ \hat{g}_{mn}= \hat{b}_{mn} + \chi_{\underline{a}}{}^{\underline{b}}\ \hat{e}_{m}{}^{\underline{a}}\ \hat{e}_{n \underline{b}} \ ,\   
\end{eqnarray}
we obtain the following expressions for ${\cal E}_{m}{}^{\underline a}$ and ${\cal E}_{m}{}^{\overline a}$
\begin{eqnarray}
{\cal E}_{m}{}^{\underline a} &=&  -\frac{1}{\sqrt2} \left( 
\hat b_{pm} \ \hat{e}^{p\underline a} + \chi^{\underline a}{}_{\underline b} \  \hat{e}_{m}{}^{\underline b}\right)\ , \cr
{\cal E}_{m}{}^{\overline a} &=&  \frac{1}{\sqrt2} \left( 
\hat b_{mp} (\chi^{-\frac12})^{\underline b}{}_{\underline c}\ \hat{e}^{p\underline c} + (\chi^{\frac12})^{\underline b}{}_{\underline c} \  \hat{e}_{m}{}^{\underline c}\right)\ \delta_{\underline b}{}^{\overline{a}} \ .
\end{eqnarray}

while ${\cal E}^{m}{}^{\underline a}$ and ${\cal E}^{m}{}^{\overline a}$ are
\begin{eqnarray}
{\cal E}^{m}{}^{\underline a} &=&  \frac{1}{\sqrt2} \ \hat{e}^{m}{}^{\underline a}\ , \cr
{\cal E}^{m}{}^{\overline a} &=& \delta^{\overline a}{}_{\underline c} \ \frac{1}{\sqrt2}  \ (\chi^{-\frac12})^{\underline c}{}_{\underline b} \  \hat{e}^{m\underline b} \ .
\end{eqnarray}

Similarly, using ${\cal E}_{{\cal N}\overline{\alpha}} \, {\cal E}^{{\cal N}}{}_{\underline{b}}=0$ and the gauge-fixing condition ${\cal E}_{{\cal \tilde\nu}}{}^{\overline{a}}=0$, we find
\begin{equation}
\hat{A}_{m\overline{\alpha}} = \frac{1}{\sqrt{2}} \, \hat{A}_{m}{}_{\tilde\mu} \, \hat{A}_{n}{}^{\tilde\mu} \, \hat{e}^{n}{}_{\overline{\alpha}} .
\end{equation}

As a consequence, the vielbein component takes the more compact form
\begin{equation}
{\cal E}_{m}{}^{\overline{\alpha}} = \frac{1}{\sqrt{2}} \left(\hat{c}_{mp} + \hat{g}_{mp}\right) \hat{e}^{p\overline{\alpha}} .
\end{equation}

Our next objective is to express the components $\hat{e}^{m}{}_{\overline{\alpha}}$, $\hat{e}^{\tilde\mu}{}_{\overline{a}}$, and $\hat{e}^{\tilde\mu}{}_{\overline{\alpha}}$ in terms of the supergravity {\it dof} $\hat{e}^{m}{}_{\underline{a}}$ and $\hat{b}_{mn}$ and the gauge objects $e_{\tilde\mu}{}^{\overline{\alpha}}$ and $\hat{A}_{m}{}^{\tilde\mu}$, or equivalently ${\cal E}_{\tilde\mu}{}^{\underline{a}}$.

We start with the condition ${\cal E}_{\tilde\mu}{}^{\overline{a}}=0$, implying
\begin{equation}
\hat{e}_{\tilde\mu}{}^{\overline{a}} = \frac{1}{\sqrt{2}}\hat{A}_{p\tilde\mu} \hat{e}^{p\overline{a}}.
\end{equation}
Using equation (\ref{vielbeins}), this becomes
\begin{equation}
\hat{e}_{\tilde\mu}{}^{\overline{a}}=\frac{1}{\sqrt{2}}\hat{A}_{p\tilde\mu} \ \delta^{\overline{a}}_{\underline{c}}\ (\chi^{-\frac12})^{\underline{c}}{}_{\underline{d}} \ \hat{e}^{p\underline{d}}. \label{emua}
\end{equation}

To find an expression for $\hat{e}^{n}{}_{\overline{\alpha}}$, we use the identity from equation (\ref{Id3}), $\hat{e}_{n}{}^{\overline{b}}\ \hat{e}^{n}{}_{\overline{\beta}}= - \hat{e}_{\tilde\mu}{}^{\overline{b}} \ \hat{e}^{\tilde\mu}{}_{\overline{\beta}}$, which yields
\begin{equation}
\hat{e}^{n}{}_{\overline{\beta}} = - \hat{e}^{n}{}_{\underline{a}} \ (\chi^{\frac12})^{\underline{a}}{}_{\underline{c}}\ \delta^{\underline{c}}_{\overline{b}}\ \hat{e}_{\tilde\mu}{}^{\overline{b}}\ \hat{e}^{\tilde\mu}{}_{\overline{\beta}}.
\end{equation}
Substituting the result from equation (\ref{emua}) leads to the simplified expression
\begin{eqnarray}
\hat{e}^{n}{}_{\overline{\beta}} &=& - \frac{1}{\sqrt{2}} \ \hat{e}^{\tilde\mu}{}_{\overline{\beta}} \hat{A}^{n}{}_{\tilde\mu}. \label{enbeta}
\end{eqnarray}

We now compute ${\cal E}_{\tilde\mu}{}^{\overline{\alpha}}$:
\begin{eqnarray}
{\cal E}_{\tilde\mu}{}^{\overline{\alpha}}&=&
\hat{e}_{\tilde\mu}{}^{\overline{\alpha}} -\frac{1}{\sqrt2}\ \hat{A}_{p\tilde\mu} \ \hat{e}^{p\overline{\alpha}}\cr
&=& \left(\kappa_{\tilde\mu\tilde\nu}+ \frac12 \hat{A}_{p\tilde\mu} \hat{A}^{p}{}_{\tilde\nu} \right)\ \hat{e}^{\nu\overline{\alpha}}\cr
&=& \Box_{\tilde\mu\tilde\nu} \ \hat{e}^{\tilde\nu\overline{\alpha}}\; ,
\end{eqnarray}
where the second line follows from using equation (\ref{enbeta}), and the third line uses the identity $\frac{1}{2} \hat{A}_{p\tilde\mu} \hat{A}^{p}{}_{\tilde\nu} = \hat{\cal{E}}_{\tilde\mu\underline{b}} \hat{\cal{E}}_{\tilde\nu}{}^{\underline b}$.

In the DFT frame, this component is
\begin{equation}
{\cal E}_{\tilde\mu}{}^{\overline{\alpha}} = (\Box^{\frac12})_{\tilde\mu\tilde\nu} \ e^{\nu\overline{\alpha}},
\end{equation}
where $e^{\tilde\nu\overline{\alpha}}$ is the (constant, after gauge fixing) vielbein of the Killing metric, $\kappa_{\tilde\mu\tilde\nu} = e_{\tilde\mu}{}^{\overline{\alpha}}\ \kappa_{\overline{\alpha\beta}} \ e_{\tilde\nu}{}^{\overline{\beta}}$, which must not be confused with $\hat{e}^{\nu\overline{\alpha}}$ satisfying (\ref{CF}). Hence we get
\begin{eqnarray}
\hat{e}_{\tilde\mu}{}^{\overline{\alpha}}=(\Box^{-\frac12})_{\tilde\mu\tilde\nu} \ e^{\tilde\nu\overline{\alpha}}    
\end{eqnarray}
and
\begin{eqnarray}
\hat{e}^{n}{}_{\overline{\beta}}=  - \frac{1}{\sqrt2} (\Box^{-\frac12})^{\tilde\mu}{}_{\tilde\nu} e^{\tilde\nu}{}_{\overline{\beta}}      
\hat{A}^{n}{}_{\tilde\mu} \; .
\end{eqnarray}

With all previous results, and the relation ${\cal E}_{\tilde\mu}{}^{\underline{a}}=- \frac{1}{\sqrt 2}\hat{A}_{p\tilde\mu} \hat{e}^{p \underline{a}}$  we can now write the complete set of components for ${\cal E}_{\cal N}{}^{\cal A}$ as

\begin{eqnarray}
{\cal E}^{m}{}^{\underline a} &=&  
\frac{1}{\sqrt2} \ \hat{e}^{m}{}^{\underline a}\ , 
\;\;\;\;\;\;\;\;\;\;\;\;\;\;\;\;\;\;\;\;\;\;\;\;\;\;\;\;\;\;\;\;\;\;\;\;\;\;\;
{\cal E}_{m}{}^{\underline a} =  
\frac{1}{\sqrt2} \ \left(\hat{b}_{mp} \ \hat{e}^{p\underline{a}} - \chi^{\underline{a}}{}_{\underline{b}} \ \hat{e}_{m}{}^{\underline{b}}\right) \ , \cr
{\cal E}^{m}{}^{\overline a} &=& 
 \frac{1}{\sqrt2}  \ (\chi^{-\frac12})^{\underline c}{}_{\underline b} \  \hat{e}^{m\underline b} \ \delta^{\overline a}{}_{\underline c}\ ,
 \;\;\;\;\;\;\;\;\;\;\;\;\;
{\cal E}_{m}{}^{\overline a} = 
\frac{1}{\sqrt2} \ \left(\hat{b}_{mp} (\chi^{-\frac12})^{\underline{b}}{}_{\underline{c}}\ \hat{e}^{p\underline{c}} + (\chi^{\frac12})^{\underline{b}}{}_{\underline{c}} \ \hat{e}_{m}{}^{\underline{c}}\right)\ \delta^{\overline{a}}_{\underline{b}} \ ,\cr
{\cal E}^{m}{}^{\overline \alpha} &=& 
\frac{1}{\sqrt2} \hat{e}^{m}{}_{\underline{c}} (\chi^{-\frac12})^{\underline{c}}{}_{\underline{d}} \; {\cal E}^{\tilde\nu\underline{d}} \; e_{\tilde\nu}{}^{\overline{\alpha}},
 \;\;\;\;\;\;\;
{\cal E}_{m}{}^{\overline \alpha} = 
-\frac{1}{\sqrt2} \ \left(\hat{b}_{mp} (\chi^{-\frac12})_{\underline{a}}{}^{\underline{b}}\ \hat{e}^{p\underline{a}} - (\chi^{\frac12})_{\underline{a}}{}^{\underline{b}} \ \hat{e}_{m}{}^{\underline{a}}\right)\ {\cal E}^{\tilde\nu}{}_{\underline{b}} \; e_{\tilde\nu}{}^{\overline{\alpha}}  \ ,\cr
{\cal E}_{\tilde\mu}{}^{\underline a} &=&  
{\cal E}_{\tilde\mu}{}^{\underline a}\ , 
\;\;\;\;\;\;\;\;\;\;\;\;\;\;\;\;\;\;\;\;\;\;\;\;\;\;\;\;\;
{\cal E}_{\tilde\mu}{}^{\overline a} = 0 \ ,
\;\;\;\;\;\;\;\;\;\;\;\;\;\;\;\;\;\;\;\;
{\cal E}_{\tilde\mu}{}^{\overline \alpha} = 
(\Box^{\frac12})_{\tilde\mu\tilde\nu} \ e^{\tilde\nu\overline{\alpha}}  \ .\label{parametrization}
\end{eqnarray}

\subsection{Generalized fluxes}
Turning now to the generalized fluxes in the SUGRA framework, the computations of Appendix \ref{AppB} yield
\begin{eqnarray}
{\cal F}_{\underline a \overline{bc}} &=&
\delta_{\overline{bc}}^{\underline{ef}} \ 
{\cal F}^{(+)}_{\underline{aef}} \; ,\cr
{\cal F}_{\underline a \overline{\alpha\beta}} &=& 
  e^{\tilde\mu}{}_{[\overline{\alpha}}
 e^{\tilde\nu}{}_{\overline{\beta}]} \left\{\vphantom{\frac12}\right. g f_{\tilde\rho\tilde\lambda\tilde\sigma} {\cal E}^{\tilde\rho}{}_{\underline{a}} \ (\Box^{\frac12})^{\tilde\lambda}{}_{\tilde\mu} \ (\Box^{\frac12})^{\tilde\sigma}{}_{\tilde\nu}\cr 
 && \;\;\;\;\;\;\;\;\;
 + \;  {\cal E}_{\tilde\nu}{}^{\underline{f}} \
\left[ {\cal E}_{\tilde\mu}{}^{\underline{e}} \
{\cal F}^{(+)}_{\underline{aef}} 
-2 \ {\cal G}_{\underline{af}\tilde\mu}
\right]
-
\frac{1}{\sqrt{2}} \left(
D_{\underline{a}}{\cal E}_{\tilde\mu}{}^{\underline{e}} \ 
{\cal E}_{\tilde\nu}{}_{\underline{e}} 
+ D_{\underline{a}}(\Box^{\frac12})_{\tilde\mu}{}^{\tilde\rho} \
(\Box^{\frac12})_{\tilde\nu\tilde\rho}
 \right) \left.\vphantom{\frac12}\right\} \; ,
 \cr
{\cal F}_{\underline{a} \overline{b}\overline{\gamma}} &=& 
\left(  \ e^{\tilde\nu}{}_{\overline{\gamma}} \ \delta^{\underline{e}}_{\overline{b}}\right) 
\left[
{\cal E}_{\tilde\nu}{}^{\underline{d}} \ {\cal F}^{(+)}_{\underline{aed}} 
+{\cal G}_{\underline{ae}\tilde\nu}
\right]\,,\label{FDecomposition}
\end{eqnarray}
where
\begin{eqnarray}
{\cal G}_{\underline{ae\tilde\nu}} &=& 
\frac{1}{\sqrt{2}} (\chi^{-\frac12})_{\underline{e}}{}^{\underline{f}}\left[ {\cal E}_{\tilde\nu}{}^{\underline{d}} \ 
\ \left( D_{\underline{a}}(\chi^{\frac12})_{\underline{fd}} + 2 \omega_{[\underline{af}]}{}^{\underline{c}} (\chi^{\frac12})_{\underline{cd}}\right)
-  \ D_{\underline{f}}({\cal E}^{\tilde\nu}{}_{\underline{a}}) (\Box^{\frac12})_{\tilde\mu\tilde\nu}    \right]\cr
&=& 
\frac{1}{\sqrt{2}} (\chi^{-\frac12})_{\underline{e}}{}^{\underline{f}}\left[ {\cal E}_{\tilde\nu}{}^{\underline{d}} \ 
\ \left( D_{\underline{a}}(\chi^{\frac12})_{\underline{fd}} 
+  \omega_{\underline{af}}{}^{\underline{c}} (\chi^{\frac12})_{\underline{cd}}\right)
-
\left( \ D_{\underline{f}}({\cal E}^{\tilde\mu}{}_{\underline{a}})  \
+
\omega_{\underline{fa}}{}^{\underline{c}}
{\cal E}^{\tilde\mu}{}_{\underline{c}} \right)(\Box^{\frac12})_{\tilde\mu\tilde\nu} \right]\; ,
\end{eqnarray}
with ${\cal F}^{(+)}_{\underline{acd}}$ given by 

\begin{eqnarray}
{\cal F}^{(+)}_{\underline {abc}}&=&
\frac{1}{\sqrt{2}} (\chi^{-\frac12})_{[\underline b}{}^{\underline e} \ (\chi^{-\frac12})_{\underline c]}{}^{\underline f}\left[\vphantom{\frac12}\right.
\ \omega^{(+)}_{\underline{ae}\underline{f}}   
\cr
&&\;\;\;\;\;\;\;\;\;\;\;\;\;\;\;\;\;\;\;\;\;\;\;\;\;\;
+ \frac{1}{ g^2 X_R}\left( \vphantom{\frac12}\right.
\sqrt2 {\cal R}^{(+)}_{\underline{ae}}{}^{\overline{\cal CD}}   
-2\sqrt2 {\cal F}_{\underline{a}}{}^{\overline{\cal C}}{}_{\overline{\cal B}}
{\cal F}_{\underline{e}}{}^{\overline{\cal B}}{}^{\overline{\cal D}}
 -  D_{\underline a}  \left(
\ {\cal F}_{\underline{e}}{}^{\overline{\cal CD}}  
 \right)
\left. \vphantom{\frac12}\right) {\cal F}_{\underline{f}}{}_{\overline{\cal CD}} \cr
&&
\;\;\;\;\;\;\;\;\;\;\;\;\;\;\;\;\;\;\;\;\;\;\;\;\;\;\;\;\;\;\;
\;\;\;\;\;\;\;\;\;\;\;\;\;\;\;\;\;\;\;\;\;\;\;\;\;\;\;\;\;\;\;
\;\;\;\;\;\;\;\;\;\;\;\;\;\;\;\;\;\;\;\;\;\;\;\;\;\;\;\;\;\;\;
-  D_{\underline a} 
\left( (\chi^{\frac12})_{\underline{he}} \right)(\chi^{\frac12})_{\underline{f}}{}^{\underline{h}}
  \left.\vphantom{\frac12}\right]\; ,\cr&& \label{F+2}   
\end{eqnarray}
where the first line is ${\cal O}(1)$, the second line is ${\cal O}(\alpha')$, while the last line is ${\cal O}(\alpha'{}^2)$. Here we have introduced the torsionful spin connection
\begin{eqnarray}
\omega^{(+)}_{m\underline{ab}}=\omega_{m\underline{ab}}+\frac12 \hat{H}_{m\underline{ab}}\    
\end{eqnarray}
and we have expressed ${\cal F}^{(+)}$ in terms of the generalized Riemann tensor (\ref{GenRim}) which will be formally presented in the next section.

For later reference we also display here the derivative expansion (up to ${\cal O}(\alpha'^{\frac32})$) of the generalized fluxes. In addition we make use also of the generalized Bergshoeff-de Roo identification to present these in a convenient way

\begin{eqnarray}
{\cal F}_{\underline a \overline{bc}} &=& 
\delta_{\overline{bc}}^{\underline{ef}} \ 
\left[ \frac{1}{\sqrt{2}} {\omega}^{(+)}_{\underline{aef}} + \frac{1}{g^2 \  X_R}\left( {\cal R}^{(+)}_{\underline{a}\underline{e}}{}^{\overline{\cal CD}} - {\cal T}_{\underline{a}\underline{e}}{}^{\overline{\cal CD}} \right)
{\cal F}_{\underline{f}\overline{\cal CD}}\right]
 +{\cal O}(\alpha'{}^2)\; ,\cr
{\cal F}_{\underline a \overline{\alpha\beta}} &=& 
 \frac{1}{X_R} e^{\tilde\mu}{}_{[\overline{\alpha}}
 e^{\tilde\nu}{}_{\overline{\beta}]} 
 (t_{\tilde\mu})^{\overline{\cal CD}} (t_{\tilde\nu})^{\overline{\cal EF}}
 \left[\vphantom{\frac12}\right. 2\ \eta_{\overline{\cal FD}} {\cal F}_{\underline{a} \overline{\cal CE} } 
 + \frac{1}{g^2 \ X_R} \left(-{\cal R}{}^{(+)}_{\underline{ae}\overline{\cal CD}} + {\cal T}_{\underline{ae}\overline{\cal CD}}\right) {\cal F}^{\underline{e}}{}_{\overline{\cal EF}}
 \left.\vphantom{\frac12}\right] +{\cal O}(\alpha'{}^2)\; ,
 \cr
{\cal F}_{\underline{a} \overline{b}\overline{\gamma}} &=& 
\delta^{\underline{e}}_{\overline{b}} e^{\tilde\mu}{}_{\overline{\gamma}}
 (t_{\tilde\mu})^{\overline{\cal CD}} 
 \frac{1}{g \ X_R} 
 \left(-\frac12 {\cal R}{}^{(+)}_{\underline{ae}\overline{\cal CD}} + {\cal T}_{\underline{ae}\overline{\cal CD}}\right) 
 +{\cal O}(\alpha'{}^{\frac32})\; ,\label{FluxExpansion}
\end{eqnarray}
here we have introduced the tensor
\begin{eqnarray}
{\cal T}_{\underline{ab}}{}^{\overline{\cal CD}} =
\frac{1}{\sqrt2} \left( 
D_{\underline{a}}{\cal F}_{\underline{b}}{}^{\overline{\cal{CD}}} 
+ \omega^{(+)}_{\underline{ab}}{}^{\underline{e}} {\cal F}_{\underline{e}}{}^{\overline{\cal{CD}}} 
- 2 \sqrt{2} \ {\cal F}_{\underline{a}}{}^{[\overline{\cal C}}{}_{\overline{\cal E}} \; 
{\cal F}_{\underline{b}}{}^{\overline{\cal D}]\overline{\cal E}}\right)    \;,
\end{eqnarray}
\bigskip
which satisfies the following useful identities
\begin{eqnarray}
{\cal T}_{[\underline{ab}]}{}^{\overline{\cal CD}} =
\frac14 {\cal R}^{(+)}_{\  [\underline{ab}]}{}^{\overline{\cal CD}}
- \ {\cal F}_{[\underline{a}}{}^{\overline{\cal C} \overline{\cal E}} \ 
{\cal F}_{\underline{b}]}{}^{\overline{\cal D}}{}_{\overline{\cal E}}
+\frac{\sqrt{2}}{4} \hat{H}_{\underline{ab}}{}^{\underline{e}} 
{\cal F}_{\underline{e}}{}^{\overline{\cal CD}} \;,
\end{eqnarray}
\begin{eqnarray}
D_{[\underline{a}}{\cal T}_{\underline{b}]\underline{e}}{}^{\overline{\cal CD}}
+ \omega_{[\underline{ab}]}{}^{\underline{f}} 
{\cal T}_{\underline{fe}}{}^{\overline{\cal CD}} 
+ \omega^{(+)}_{[\underline{a}|\underline{e}|}{}^{\underline{f}} 
{\cal T}_{\underline{b}]\underline{f}}{}^{\overline{\cal CD}}
- 2\sqrt{2} {\cal F}_{[\underline{a}}{}^{\overline{\cal E}[\overline{\cal C}}
{\cal T}_{\underline{b}]\underline{e}\overline{\cal E}}{}^{\overline{\cal D}]}
=\frac{\sqrt{2}}{4}
\left(
R^{(+)}_{\underline{abe}}{}^{\underline{c}} \ 
{\cal F}_{\underline{c}}{}^{\overline{\cal CD}} 
- 2 {\cal R}^{(+)}_{\underline{ab}}{}^{\overline{\overline{\cal E}[\cal C}} \ 
{\cal F}_{\underline{e}\overline{\cal E}}{}^{\overline{\cal D}]}
\right)\;.\cr\label{DTid}
\end{eqnarray}

\subsection{The generalized Riemann tensor and the action}
The gauged DFT, after explicit resolution of the strong constraint in terms of the SUGRA section (\ref{SUGRAsection}) is
\begin{eqnarray}
{\cal L}= \sqrt{-\hat{g}}\ e^{-2\hat{\phi}}  \left( \hat{R} - 4\ (\nabla \hat\phi)^2 + 4\  \Box\hat\phi -\frac{1}{12} \hat{H}_{mnp} \hat{H}^{mnp} 
-\frac{1}{4}
{F}_{mn}{}^{\tilde\mu} \ {F}^{mn}{}_{\tilde\mu}
\right)  \; ,
\end{eqnarray}
with the curvatures defined as in (\ref{HSugra}) and (\ref{Fmn}). 
We define a generalization of the Riemann tensor with torsion ${\cal R}^{(+)}_{mn}{}^{\overline{\cal CD}}$ in terms of the gauge curvature 
\begin{eqnarray}
\hat{F}_{mn\tilde\mu}=-\frac{1}{g X_R} (t_{\tilde\mu})^{\overline{\cal CD}} {\cal R}^{(+)}_{mn \overline{\cal CD}}\ .
\end{eqnarray}
After gBdR identification, it reads
\begin{eqnarray}
{\cal R}^{(+)}_{mn}{}^{\overline{\cal CD}} =
2\sqrt{2}\left( \partial_{[m}{\cal F}_{n]}{}^{\overline{\cal CD}} 
- \sqrt{2} \ {\cal F}_{[m}{}^{[\overline{\cal C}}{}_{\overline{\cal E}} \; 
{\cal F}_{n]}{}^{\overline{\cal D}]\overline{\cal E}}\right)\ ,
\end{eqnarray}
where ${\cal F}_{m}{}^{\overline{CD}}= \hat{e}_{m}{}^{\underline{a}}\ {\cal F}_{\underline{a}}{}^{\overline{CD}}$. The Lagrangian then reads 
\begin{eqnarray}
{\cal L}= \sqrt{-\hat{g}}\ e^{-2\hat{\phi}}  \left( \hat{R} - 4\ (\nabla \hat\phi)^2 + 4\  \Box\hat\phi -\frac{1}{12} \hat{H}_{mnp} \hat{H}^{mnp} 
-\frac{1}{4}\frac{1}{g^2 X_R} 
{\cal R}^{(+)}_{mn}{}^{\overline{\cal CD}} \ 
{\cal R}^{(+)}{}^{mn}{}_{\overline{\cal CD}}
\right)  \; .
\end{eqnarray}

Notice that the leading order contribution of the generalized Riemann tensor when $\overline{\cal CD}$ is evaluated in $\overline{cd}$ is
\begin{eqnarray}
{\cal R}^{(+)}_{\underline{ab}}{}^{\overline{cd}} &=&
 2\sqrt{2}\left( 
D_{[\underline{a}}{\cal F}_{\underline{b}]}{}^{\overline{cd}} 
+ \omega_{[\underline{ab}]}{}^{\underline{e}} {\cal F}_{\underline{e}}{}^{\overline{cd}} 
- \sqrt{2} \ {\cal F}_{[\underline{a}}{}^{[\overline{c}}{}_{\overline{\cal E}} \; 
{\cal F}_{\underline{b}]}{}^{\overline{d}]\overline{\cal E}}\right)\cr \cr
&=&
 2\sqrt{2} \delta^{\overline{cd}}_{\underline{ef}}  \left( \vphantom{\frac12}
D_{[\underline{a}}{\cal F}^{(+)}_{\underline{b}]}{}^{\underline{ef}} 
+ \omega_{[\underline{ab}]}{}^{\underline{g}} {\cal F}^{(+)}_{\underline{g}}{}^{\underline{ef}} 
+ \sqrt{2} \ \chi_{\underline{gh}} {\cal F}^{(+)}_{[\underline{a}}{}^{\underline{e g}}\; 
{\cal F}^{(+)}_{\underline{b}]}{}^{\underline{fh}}\right)\cr \cr
&=& \delta^{\overline{cd}}_{\underline{ef}} 
\ R^{(+)}_{\underline{ab}}{}^{\underline{ef}}
 \ + {\cal O}(\alpha')
\end{eqnarray}
where 
\begin{eqnarray}
 R^{(+)}_{\underline{ab}}{}^{\underline{ef}} &=& 
 2\left( 
D_{[\underline{a}}{\omega}^{(+)}_{\underline{b}]}{}^{\underline{ef}} 
+ \omega_{[\underline{ab}]}{}^{\underline{g}} {\omega}^{(+)}_{\underline{g}}{}^{\underline{ef}} 
-  \ {\omega}^{(+)}_{[\underline{a}}{}^{[\underline{e}}{}_{\underline{g}} \; 
{\omega}^{(+)}_{\underline{b}]}{}^{\underline{f}]\underline{g}}\right)\  \cr
&=& R_{\underline{ab}}{}^{\underline{ef}} 
+ \nabla_{[\underline{a}} \hat{H}_{\underline{b}]}{}^{\underline{ef}}
-\frac12 {\hat H}_{[\underline{a}}{}^{\underline{eg}} \ 
\hat{H}_{\underline{b}]}{}^{\underline{f}}{}_{\underline{g}}
\label{Riemann+}
\end{eqnarray}
is the Riemann tensor associated with the torsionful spin connection $\omega^{(+)}_{m}{}^{\underline{ab}}$. Whose torsion, $\hat{H}_{m\underline{bc}}= \hat{e}^{n}{}_{\underline{b}} \hat{e}^{p}{}_{\underline{c}}\ \hat{H}_{mnp}$ is written in terms of the generlaized Riemann tensor. $\hat{H}=d\hat{b}-\hat\Omega_{CS}$, with the Lorentz Chern-Simons term $\hat\Omega_{CS}$ obtained by imposing the gBdR identification on $\hat\Omega_{CS}(A)$ in Eq. (\ref{OCS})
\begin{eqnarray}
(\hat\Omega_{CS})_{\underline{abc}}= 
\frac{\sqrt{2}}{g^2 X_R}{\cal F}_{[\underline{a}}{}^{\overline{\cal{CD}}}  
\left( 
\frac{3}{2} {\cal R}^{(+)}_{\underline{bc}]\overline{\cal CD}} 
+ 2 {\cal F}_{\underline{b}\overline{\cal CB}}
{\cal F}_{\underline{c}]\overline{\cal D}}{}^{\overline{\cal B}}
\right)\; .\label{OmegaCS}
\end{eqnarray}

The local transformations of ${\cal R}^{(+)}_{mn}{}^{\overline{\cal CD}}$ follow readily from those of $F_{mn\tilde\mu}$ in (\ref{deltaF}):

\begin{eqnarray}
\delta {\cal R}^{(+)}_{mn}{}^{\overline{\cal CD}} &=& - g \delta F_{mn\tilde\mu} (t^{\tilde\mu})^{\overline{\cal CD}}\cr
&=& - g {\cal L}_{\xi^{p}}F_{mn\tilde\mu} (t^{\tilde\mu})^{\overline{\cal CD}}
+ g^2  f_{\tilde{\nu}\tilde\rho\tilde{\mu}} F_{mn}{}^{\tilde\rho} (t^{\tilde\mu})^{\overline{\cal CD}} \xi^{\tilde\nu}\cr
&=& {\cal L}_{\xi^{p}} {\cal R}^{(+)}_{mn}{}^{\overline{\cal CD}} + 
2 \Gamma^{[\overline{\cal{C}|E|}}  {\cal R}^{(+)}_{mn}{}_{\overline{\cal E}}{}^{\overline{\cal D}]}
\end{eqnarray}
where ${\cal L}_{\xi^{p}}$ denotes the ordinary Lie derivative, and we have used $f_{\tilde{\nu}\tilde\rho\tilde{\mu}}  (t^{\tilde\mu})^{\overline{\cal CD}} = 2 (t_{\tilde\nu})^{[\overline{\cal C}|\overline{\cal E}|} (t_{\tilde\rho})_{\overline{\cal E}}{}^{\overline{\cal D}]}$.

Hence, ${\cal R}^{(+)}_{mn}{}^{\overline{\cal CD}}$ transforms in a mixed fashion: it behaves as a 2-form under diffeomorphisms and covariantly under Lorentz transformations in the extended space, once the gBdR identification is implemented.

When explicitly expressed in terms of derivatives of the Lorentz-covariant degrees of freedom, this extended Lorentz transformation induces a Green-Schwarz-like transformation. Prior to performing this derivative expansion, however, the action remains manifestly Lorentz invariant.

\section{Derivative expansion}

In this section, we present the derivative expansion of the action obtained in the previous section up to $\mathcal{O}(\alpha'^2)$. Expanding the generalized fluxes as in (\ref{FluxExpansion}) and making use of identity (\ref{DTid}), we find, after some algebra,
\newpage

\begin{eqnarray}
{\cal R}^{(+)}_{\underline{ab}}{}_{\overline{cd}} &=&
 \delta^{\underline{ef}}_{\overline{cd}}\left(
 R^{(+)}_{\underline{abef}} + 
\frac{1}{g^2 X_R} \Delta R^{(+)}_{\underline{abef}}
\right) + {\cal O}(\alpha'{}^2)\;,\cr
{\cal R}^{(+)}_{\underline{ab}\ \overline{\beta\gamma}} &=&
 \frac{1}{X_R} e^{\tilde\mu}{}_{[\overline{\beta}}e^{\tilde\nu}{}_{\overline{\gamma}]}  (t_{\tilde\mu})^{\overline{\cal CD}} (t_{\tilde\nu})^{\overline{\cal EF}} 
 \left( 
-2 \eta_{\overline{\cal DE}}\ {\cal R}^{(+)}_{\underline{ab}\overline{\cal CF}} +
\frac{1}{g^2 X_R} \Delta{\cal R}_{\underline{ab}\overline{\cal CD EF}}
\right) + {\cal O}(\alpha'{}^2)\;,\cr
{\cal R}^{(+)}_{\underline{ab}\ \overline{c\gamma}} &=&
 \frac{1}{g X_R} e^{\tilde\mu}{}_{\overline \gamma} \delta^{\underline{e}}_{\overline{c}} (t_{\tilde\mu})^{\overline{\cal CD}} 
 \Delta {\cal R}^{(+)}_{\underline{abe}\overline{\cal CD}}
 + {\cal O}(\alpha'{}^{\frac32})\;,\label{R+Expansion}
\end{eqnarray}
with
\begin{eqnarray}
\Delta R^{(+)}_{\underline{abef}}&=&
2\sqrt{2}\  \hat{\cal D}^{(+)}_{[\underline{a}} 
{\cal R}^{(+)}_{\underline{b}][\underline{e}}{}^{\overline{\cal CD}}
{\cal F}_{\underline{f}]\overline{\cal CD}} 
- {\cal R}^{(+)}_{[\underline{a}|\underline{e}|}{}^{\overline{\cal CD}}\ 
{\cal R}^{(+)}_{\underline{b}]\underline{f}}{}_{\overline{\cal CD}}
- \left({R}^{(+)}_{\underline{ab}[\underline{e}|}{}^{\underline{g}} {\cal F}_{\underline{g}}{}^{\overline{\cal CD}} 
- 2 {\cal R}^{(+)}_{\underline{ab}}{}^{\overline{\cal CG}} 
{\cal F}_{[\underline{e}}{}^{\overline{\cal D}}{}_{|\overline{\cal G}|}\right)
{\cal F}_{\underline{f}]\overline{\cal CD}}\ , 
\cr
\cr 
\Delta{\cal R}_{\underline{ab}\overline{\cal CD EF}}&=& 
 -2 \sqrt{2} \hat{\cal D}^{(+)}_{[\underline{a}} {\cal R}{}^{(+)}_{\underline{b}]\underline{e}}{}_{\overline{\cal CD}} \ 
 {\cal F}^{\underline{e}}{}_{\underline{\cal EF}}
 + {\cal R}^{(+)}_{[\underline{a}|\underline{e}\overline{\cal CD}|} \ 
 {\cal R}^{(+)}_{\underline{b}]}{}^{\underline{e}}{}_{\overline{\cal EF}}
 +\left(R^{(+)}_{\underline{abe}}{}^{\underline{g}}\ 
 {\cal F}_{\underline{g}\overline{\cal CD}}
 - 2 {\cal R}^{(+)}_{\underline{ab}[\overline{\cal C}}{}^{\overline{\cal G}} \
 {\cal F}_{|\underline{e}|\overline{\cal D}]\overline{\cal G}} \right)
{\cal F}^{\underline{e}}{}_{\overline{\cal EF}} \, ,
\cr
\cr 
\Delta {\cal R}^{(+)}_{\underline{abe}}{}^{\overline{\cal CD}}&=& 
- \sqrt{2} 
\hat{\cal D}^{(+)}_{[\underline{a}} {\cal R}{}^{(+)}_{\underline{b}]\underline{e}}{}^{\overline{\cal CD}} 
- R^{(+)}_{\underline{abef}}{\cal F}^{\underline{f}\overline{\cal CD}}
 -2 \ {\cal R}^{(+)}_{\underline{ab}}{}^{\overline{\cal E}[\overline{\cal C}}
 {\cal F}_{\underline{e}\overline{\cal E}}{}^{\overline{\cal D}]}\ ,\label{DR+Expansion}
\end{eqnarray} 
where we have introduced
\begin{eqnarray}
\hat{\cal D}^{(+)}_{\underline{a}} 
{\cal R}^{(+)}_{\underline{b}\underline{e}}{}^{\overline{\cal CD}} 
=
D_{\underline{a}} {\cal R}{}^{(+)}_{\underline{b}\underline{e}}{}^{\overline{\cal CD}} 
+ \omega_{\underline{ab}}{}^{\underline{f}} 
 {\cal R}{}^{(+)}_{\underline{fe}}{}^{\overline{\cal CD}} 
 + \omega^{(+)}_{\underline{a}\underline{e}}{}^{\underline{f}} 
 {\cal R}{}^{(+)}_{\underline{b}\underline{f}}{}^{\overline{\cal CD}} 
 + 2 \sqrt{2} {\cal F}_{\underline{a}}{}^{[\overline{\cal C}}{}_{\overline{\cal E}} \ 
 {\cal R}^{(+)}_{\underline{b}\underline{e}}{}^{|\overline{\cal E}|\overline{\cal D}]} \;.
\end{eqnarray}
It is worth noticing that $\hat{\cal D}^{(+)}_{\underline{a}}$ is a differential operator that changes the symmetry properties after acting on $R^{(+)}$: ${\cal R}^{(+)}_{\underline{b}\underline{e}}{}^{\overline{\cal CD}} =
- {\cal R}^{(+)}_{\underline{e}\underline{b}}{}^{\overline{\cal CD}}$ while 
$\hat{\cal D}^{(+)}_{\underline{a}} 
{\cal R}^{(+)}_{\underline{b}\underline{e}}{}^{\overline{\cal CD}} \neq
- \hat{\cal D}^{(+)}_{\underline{a}} 
{\cal R}^{(+)}_{\underline{e}\underline{b}}{}^{\overline{\cal CD}} $. 

Then
\begin{eqnarray}
\frac{1}{g^2 X_R}{\cal R}^{(+)}_{\underline{ab}\overline{cd}} 
{\cal R}^{(+)}{}^{\underline{ab}\overline{cd}}   &=& 
\frac{1}{g^2 X_R}\left(\vphantom{\frac12}\right.
R^{(+)}_{\underline{abef}} 
R^{(+)}{}^{\underline{abef}} +
\frac{2}{g^2\ X_R}  R^{(+)}_{\underline{abef}} 
\Delta R^{(+)}{}^{\underline{abef}} 
\left.\vphantom{\frac12}\right) +{\cal O}(\alpha'{}^3)\;,
\cr
\cr
\frac{1}{g^2 X_R}{\cal R}^{(+)}_{\underline{ab}\overline{\beta\gamma}} 
{\cal R}^{(+)}{}^{\underline{ab}\overline{\beta\gamma}}   &=&  
\frac{1}{g^2 X_R}\left(\vphantom{\frac12}\right.
(\delta_{\overline{\cal E}}{}^{\overline{\cal E}}-2) 
{\cal R}^{(+)}_{\underline{ab}\overline{\cal CD}} 
{\cal R}^{(+)}{}^{\underline{ab}\overline{\cal CD}} 
- \frac{4}{g^2 X_R} {\cal R}^{(+)}_{\underline{ab}\overline{\cal C}}{}^{\overline{\cal D}} 
\Delta{\cal R}^{(+)}{}^{\underline{ab}\overline{\cal CE}}{}_{\overline{\cal ED}} 
\left.\vphantom{\frac12}\right) +{\cal O}(\alpha'{}^3)\;,
\cr
\cr
\frac{2}{g^2 X_R}{\cal R}^{(+)}_{\underline{ab}\overline{c\gamma}} 
{\cal R}^{(+)}{}^{\underline{ab}\overline{c\gamma} }  &=& 
\frac{1}{g^2 X_R} \frac{2}{g^2 X_R} \left(\vphantom{\frac12}\right.
- \ \Delta{\cal R}^{(+)}_{\underline{abe}\overline{\cal CD}} 
\Delta{\cal R}^{(+)}{}^{\underline{abe}\overline{\cal CD}}
\left.\vphantom{\frac12}\right) +{\cal O}(\alpha'{}^3)\;,
\end{eqnarray}
where we used $g^{-2}\sim\alpha'$. The infinite trace is regularized in terms of $X_R$
\begin{eqnarray}
(\delta_{\overline{\cal E}}{}^{\overline{\cal E}}-2)= X_R^{-1} \;.  \label{XR}  
\end{eqnarray}
Indeed, this follows from consistency relations with the algebra and the normalization of the generators (\ref{normalization})
\begin{eqnarray}
\kappa_{\overline{\alpha\beta}}=-f_{\overline{\alpha\gamma}}{}^{\overline{\delta}}\ f_{\overline{\beta\delta}}{}^{\overline{\gamma}} =
- \frac{4}{X_R} (t_{\overline{\gamma}})^{\overline{\cal A}}{}_{[\overline{\cal C}}
 (t_{\overline{\alpha}})_{\overline{\cal D}]}{}_{\overline{\cal A}}
  (t_{\overline{\beta}})^{\overline{\cal C B}}
   (t^{\overline{\gamma}})_{\overline{\cal B}}{}^{\overline{\cal D}}
   = X_R \left(\delta^{\overline{\cal A}}_{\overline{\cal A}} - 2\right) \ \kappa_{\overline{\alpha\beta}}\; .
\end{eqnarray}
Then
\begin{eqnarray}
\frac{1}{g^2 X_R}{\cal R}^{(+)}_{\underline{ab}\overline{\cal CD}} 
{\cal R}^{(+)}{}^{\underline{ab}\overline{\cal CD}}   &=& 
\frac{1}{g^2 X_R} \frac{1}{X_R} {\cal R}^{(+)}_{\underline{ab}\overline{\cal CD}} 
{\cal R}^{(+)}{}^{\underline{ab}\overline{\cal CD}} 
\cr
&+&  \frac{2}{g^2 X_R}\left(\vphantom{\frac12}\right.
R^{(+)}_{\underline{abef}} 
R^{(+)}{}^{\underline{abef}} +
\frac{1}{g^2 X_R} R^{(+)}_{\underline{abef}} 
\Delta R^{(+)}{}^{\underline{abef}} 
-\frac{4}{g^2 X_R} {\cal R}^{(+)}_{\underline{ab}\overline{\cal C}}{}^{\overline{\cal D}} 
\Delta{\cal R}^{(+)}{}^{\underline{ab}[\overline{\cal CE}]}{}_{[\overline{\cal ED}]} \cr
&&-
\frac{2}{g^2 X_R} 
\Delta{\cal R}^{(+)}_{\underline{abe}\overline{\cal CD}} 
\Delta{         \cal R}^{(+)}{}^{\underline{abe}\overline{\cal CD}}
\left.\vphantom{\frac12}\right) +{\cal O}(\alpha'{}^3)\;.
\end{eqnarray}
The first term in the $rhs$ combines with the $lhs$ producing
\begin{eqnarray}
\frac{1}{g^2 X_R}{\cal R}^{(+)}_{\underline{ab}\overline{\cal CD}} 
{\cal R}^{(+)}{}^{\underline{ab}\overline{\cal CD}}   &=& 
\frac{b}{2} \left(\vphantom{\frac12}\right.
R^{(+)}_{\underline{abef}} 
R^{(+)}{}^{\underline{abef}} +
\frac{2}{g^2 X_R} R^{(+)}_{\underline{abef}} 
\Delta R^{(+)}{}^{\underline{abef}} 
- \frac{4}{g^2 X_R} {\cal R}^{(+)}_{\underline{ab}\overline{\cal C}}{}^{\overline{\cal D}} 
\Delta{\cal R}^{(+)}{}^{\underline{ab}[\overline{\cal CE}]}{}_{[\overline{\cal ED}]} \cr
&&-
\frac{2}{g^2 X_R} 
\Delta{\cal R}^{(+)}_{\underline{abe}\overline{\cal CD}} 
\Delta{\cal R}^{(+)}{}^{\underline{abe}\overline{\cal CD}}
\left.\vphantom{\frac12}\right) +{\cal O}(\alpha'{}^3)\;,
\end{eqnarray}
where we have introduced the parameter $b$ as
\begin{eqnarray}
b=\frac{2}{g^2(-1+X_R)}    \;.
\end{eqnarray}
We see from this expression that the apparent singularity on the $lhs$ when $k\to\infty$ or equivalently, from (\ref{XR}) $X_R\to0$ is compensated, indeed after explicitly working the Riemann squared interaction we see that the leading order is $\frac{b}{2} R^{(+)}_{\underline{abef}} R^{(+)}{}^{\underline{abef}}$ which is finite in that limit. Similar manipulations at the subleading order lead to 
\begin{eqnarray}
\frac{1}{g^2 X_R}{\cal R}^{(+)}_{\underline{ab}\overline{\cal CD}} 
{\cal R}^{(+)}{}^{\underline{ab}\overline{\cal CD}}   &=& 
\frac{b}{2}
R^{(+)}_{\underline{abef}} 
R^{(+)}{}^{\underline{abef}} \cr
&+&
b^2 \left[ \hat D^{(+)}{}^{[\underline{a}} R^{(+)}{}^{\underline{b}]\underline{ecd}}
\hat D^{(+)}_{\underline{a}} R^{(+)}_{\underline{b}\underline{ecd}} 
+R^{(+)}{}^{\underline{abef}} \left( R^{(+)}_{\underline{aced}} 
R^{(+)}_{\underline{b}}{}^{\underline{cd}}{}_{\underline{f}}
-\frac12 R^{(+)}_{\underline{ae}}{}^{\underline{cd}} R^{(+)}_{\underline{bfcd}} \right)
\right] \cr\cr
&& \;\;\;\;\;\;\;\;\;\;\;\;\;\;\;\;\;\;\;\;\;\;\;\;\;\;\;\;\;\;\;\;\;\;\;\;\;
 \;\;\;\;\;\;\;\;\;\;\;\;\;\;\;\;\;\;\;\;\;\;\;\;\;\;\;\;\;\;\;\;\;\;\;\;\;\;
 \;\;\;\;\;\;\;\;\;\;\;\;\;\;\;\;\;\;\;\;\;\;\;\;\;\;\;\;\;\;
+ {\cal O}(\alpha'{}^3)\cr \cr 
&&\label{Lalpha2}
\end{eqnarray}
which has only 3 couplings at six derivative order, that is two order of magnitude less as compared with the expressions following from the gBdR identification in the DFT frame \cite{Baron:2020xel} containing $\sim 300$ terms in the heterotic (monoparametric) case. As we anticipated, this is written in terms of physical (Lorentz covariant) tensors. Indeed, both $R^{(+)}_{\underline{abcd}}$ and $\hat{D}^{(+)}_{\underline{a}}R^{(+)}_{\underline{bcde}}$ are Lorentz covariant, as they are defined combining Lorentz covariant tensors $R_{\underline{abcd}}$ and $\hat{H}_{\underline{abc}}$ and covariant derivatives of them (see Eq. (\ref{Riemann+}) and (\ref{DR+})).

\section{Partial checks}

In this section we demonstrate that, in agreement with previous literature, there are no pure gravitational couplings of order Riemann${}^3$ at ${\cal O}(\alpha'^2)$.\footnote{Absence of $R^3$ terms was already verified in \cite{Hronek:2021nqk} in the DFT formulation. Here we will independently  prove it in the our SUGRA scheme.} The only pure gravity contributions expected at this order are the Lorentz–Chern–Simons squared term arising from $\hat{H}^2$.

The sources for pure gravity couplings at ${\cal O}(\alpha'{}^2)$ are:
\begin{eqnarray}
-\frac{1}{12} \hat{H}^2\left.\vphantom{\frac12}\right|_{b_{mn}=0}
&=&-\frac{1}{12}(\hat{\Omega}_{CS})_{\underline{abc}}
(\hat{\Omega}_{CS})^{\underline{abc}}\cr
&=& -\frac{3}{16}\ b^2 \ \hat{\Omega}^{(+)}_{\underline{abc}}
\hat{\Omega}^{(+)}{}^{\underline{abc}} + {\cal O}(\alpha'{}^3)\; ,
 \end{eqnarray}
which are the only ones expected to contribute. 

Here we used the gBdR identification on (\ref{OmegaCS}) obtaining
\begin{eqnarray}
(\hat{\Omega}_{CS})_{\underline{abc}}= \frac23 b \ \Omega^{(+)}_{\underline{abc}} + {\cal O}(\alpha'{}^2)    \;,
\end{eqnarray}
with
\begin{eqnarray}
\hat{\Omega}^{(+)}_{\underline{mnp}}=
\hat{\omega}^{(+)}_{[m}{}_{\underline{a}}{}^{\underline{b}} 
\partial_{n}\hat{\omega}^{(+)}_{p]}{}_{\underline{b}}{}^{\underline{a}}
+\frac23 \hat{\omega}^{(+)}_{[m}{}_{\underline{a}}{}^{\underline{b}} 
\hat{\omega}^{(+)}_{n}{}_{\underline{b}}{}^{\underline{c}} 
\hat{\omega}^{(+)}_{p]}{}_{\underline{c}}{}^{\underline{a}}   \;.
\end{eqnarray}

Another potential source is the $(R^{(+)}){}^2$ couplings. However, these ultimately cancel. Indeed, setting $b_{mn}=0$
\begin{eqnarray}
-\frac{b}{8}  R^{(+)}_{\underline{abcd}} \  R^{(+)}{}^{\underline{abcd}} 
\left.\vphantom{\frac12}\right|_{b_{mn}=0}
&=& -\frac{b}{8}  R_{\underline{abcd}} \  R^{\underline{abcd}} 
+ \frac{b^2}{4}  R^{\underline{abcd}} \ 
\nabla_{\underline{a}}\hat\Omega_{\underline{bcd}} + {\cal O}(\alpha'{}^3)\cr
&=& -\frac{b}{8}  R_{\underline{abcd}} \  R^{\underline{abcd}}+ {\cal O}(\alpha'{}^3)
\end{eqnarray}
due to the antisymmetry of $\hat\Omega$ and the Bianchi identity of the Riemann tensor.  

The only remaining potential pure gravitational terms are of order $b^2$. These appear as
\begin{eqnarray}
&& b^2 \left[ \hat D^{(+)}{}^{[\underline{a}} R^{(+)}{}^{\underline{b}]\underline{ecd}}
\hat D^{(+)}_{\underline{a}} R^{(+)}_{\underline{b}\underline{ecd}} 
+R^{(+)}{}^{\underline{abef}} \left( R^{(+)}_{\underline{aced}} 
R^{(+)}_{\underline{b}}{}^{\underline{cd}}{}_{\underline{f}}
-\frac12 R^{(+)}_{\underline{ae}}{}^{\underline{cd}} R^{(+)}_{\underline{bfcd}} \right)
\right]\left.\vphantom{\frac12}\right|_{b_{mn}=0}=\cr
&&= \ 
b^2 \left[ \nabla^{[\underline{a}} R^{\underline{b}]\underline{ecd}}
\nabla_{\underline{a}} R_{\underline{b}\underline{ecd}} 
+R^{\underline{abef}} \left( R_{\underline{aced}} 
R_{\underline{b}}{}^{\underline{cd}}{}_{\underline{f}}
-\frac12 R_{\underline{ae}}{}^{\underline{cd}} R_{\underline{bfcd}} \right)\right] + {\cal O}(\alpha'{}^3)\;.
\label{R3}
\end{eqnarray}

We now show that these contributions can indeed be eliminated, at the cost of introducing non-purely-gravitational couplings.

Recall that the full Lagrangian carries an overall factor $e^{-2\phi}$. Consider the first term in (\ref{R3}). Using the Bianchi identity $\nabla_{[\underline{a}}R_{\underline{bc}]\underline{de}}=0$, we obtain
\begin{eqnarray}
e^{-2\phi}\ \nabla^{[\underline{a}} R^{\underline{b}]\underline{ecd}}
\nabla_{\underline{a}} R_{\underline{b}\underline{ecd}} &=&
- \frac12 e^{-2\phi}\  \nabla^{e} R^{\underline{ab}\underline{cd}}
\nabla_{\underline{a}} R_{\underline{b}\underline{ecd}} \cr
&=&  \nabla^{e} \left[- \frac12 e^{-2\phi}\ R^{\underline{ab}\underline{cd}}
\nabla_{\underline{a}} R_{\underline{b}\underline{ecd}}\right]
- e^{-2\phi}\ \nabla^{\underline{e}}\phi \ R^{\underline{ab}\underline{cd}}
\nabla_{\underline{a}} R_{\underline{b}\underline{ecd}}\cr
&& +  \frac12 e^{-2\phi}\ R^{\underline{ab}\underline{cd}}
\nabla^{\underline{e}}\nabla_{\underline{a}} R_{\underline{b}\underline{ecd}}
\cr
&=&  \nabla^{e} \left[- \frac12 e^{-2\phi}\ R^{\underline{ab}\underline{cd}}
\nabla_{\underline{a}} R_{\underline{b}\underline{ecd}}\right]
- e^{-2\phi}\ \nabla^{\underline{e}}\phi \ R^{\underline{ab}\underline{cd}}
\nabla_{\underline{a}} R_{\underline{b}\underline{ecd}} \cr
&&+  \frac12 e^{-2\phi}\ R^{\underline{ab}\underline{cd}}
[\nabla^{\underline{e}},\nabla_{\underline{a}}] R_{\underline{b}\underline{ecd}}
 +  \frac12 e^{-2\phi}\ R^{\underline{ab}\underline{cd}}
\nabla_{\underline{a}}\nabla^{\underline{e}} R_{\underline{b}\underline{ecd}}\; .
\label{DR2}
\end{eqnarray}
The first term in the last equality is a total derivative and is hence discarded. The second term contains $\nabla\phi$, which is not a pure gravity coupling and is therefore ignored in our present analysis. We now examine the third term.
\begin{eqnarray}
\frac12 e^{-2\phi}\ R^{\underline{ab}\underline{cd}} \ [\nabla^{\underline{e}},\nabla_{\underline{a}}] R_{\underline{b}\underline{ecd}} &=& \frac12 e^{-2\phi}\ R^{\underline{ab}\underline{cd}} 
\left[R^{\underline{e}}{}_{\underline{ab}}{}^{\underline{f}} R_{\underline{fecd}}   
+ R^{\underline{e}}{}_{\underline{ae}}{}^{\underline{f}} R_{\underline{bfcd}} 
+ 2 R^{\underline{e}}{}_{\underline{ac}}{}^{\underline{f}} R_{\underline{befd}} 
\right]\cr
&=& R_{\underline{af}} \left(
\frac12 e^{-2\phi}\ R^{\underline{ab}\underline{cd}} R_{\underline{b}}{}^{\underline{f}}{}_{\underline{cd}}  \right)
+ \frac12 e^{-2\phi}\ R^{\underline{ab}\underline{cd}} 
\left[R^{\underline{e}}{}_{\underline{ab}}{}^{\underline{f}} R_{\underline{fecd}}   
+ 2 R^{\underline{e}}{}_{\underline{ac}}{}^{\underline{f}} R_{\underline{befd}} 
\right]\; .\cr&&\label{R3B}
\end{eqnarray}
The first term above is proportional to the Ricci tensor and can therefore be absorbed by a field redefinition of the metric (at the cost of introducing $\nabla\phi$ and $H$ couplings, which we are neglecting in the pure gravity analysis). The remaining Riemann${}^3$ structures in (\ref{R3B}) exactly cancel the explicit Riemann${}^3$ couplings appearing in (\ref{R3}).

After the above cancellations, the only surviving pure gravitational piece from (\ref{DR2}) is the last term, which we manipulate as follows:
\begin{eqnarray}
 \frac12 e^{-2\phi}\ R^{\underline{ab}\underline{cd}}
\nabla_{\underline{a}}\nabla^{\underline{e}} R_{\underline{becd}}
&=& 
\frac12 e^{-2\phi}\ R^{\underline{ab}\underline{cd}}
\nabla_{\underline{a}}\nabla^{\underline{e}} R_{\underline{c}\underline{dbe}}\cr
&=&-e^{-2\phi}\ R^{\underline{ab}\underline{cd}}
\nabla_{\underline{a}}\nabla_{\underline{c}} R_{\underline{d}}{}^{\underline{e}}{}_{\underline{be}}\cr
&=&
\nabla_{\underline{a}}\left[-e^{-2\phi}\ R^{\underline{ab}\underline{cd}}
\nabla_{\underline{c}} R_{\underline{db}}\right]
+
\nabla_{\underline{c}} \left[ \nabla_{\underline{a}}\left(e^{-2\phi}\ R^{\underline{ab}\underline{cd}} \right)
R_{\underline{db}} \right]\cr
&&
- R_{\underline{db}} \ \nabla_{\underline{c}}\nabla_{\underline{a}}\left(e^{-2\phi}\ R^{\underline{ab}\underline{cd}} \right)\; ,
\end{eqnarray}
where we used $R_{becd}=R_{cdbe}$ and $\nabla_{[\underline{e}} R_{\underline{cd}]\underline{b}}{}^{\underline{e}}=0$ in the first and second line, respectively. 

The first two terms in the final equality are total derivatives and are discarded, while the last term is proportional to the Ricci tensor. As before, it can be eliminated via a field redefinition of the metric, up to $\nabla\phi$ and $H$ terms that we ignore in this pure-gravity analysis.

We conclude that the only pure gravitational couplings at ${\cal O}(\alpha'{}^2)$ are indeed the Lorentz–Chern–Simons squared terms, as expected. All potential $R^3$ contributions are removed by a combination of total derivatives, Bianchi identities, and field redefinitions.

A full comparison with the standard formulation is more involved. The main subtlety lies in the fact that our $\hat{H}$, although proven to be all-order Lorentz invariant, differs from the conventional three-form $\tilde{H}$, which is defined exactly as
\begin{eqnarray}
\tilde{H}_{mnp}=db_{mnp}- \frac23 b \ \Omega^{(+)}_{\underline{abc}}  \;.   \label{tildeH}
\end{eqnarray}
This definition contains no explicit higher-derivative contributions beyond ${\cal O}(\alpha')$. In the standard approach, all higher-derivative terms arise recursively by rewriting $\tilde\omega^{(+)}=\omega+\frac12 \tilde{H}$, and then expressing $\tilde{H}$ in terms of $\Omega^{(+)}$, etc.

By contrast, our $\hat H$ defined via $\hat\Omega_{CS}$ contains explicit contributions to all orders. For instance,
\begin{eqnarray}
(\hat\Omega_{CS})_{\underline{abc}}= 
b \  (\hat\Omega_{CS}^{(+)})_{\underline{abc}}
+ b^2 \ (\hat\Omega_{CS}^{(2)})_{\underline{abc}}+ {\cal O}(\alpha'{}^3)\;.
\end{eqnarray}
The explicit form of $\hat{\Omega}^{(2)}$ is
\begin{eqnarray}
(\hat\Omega_{CS}^{(2)})_{\underline{abc}}&=&
\frac34 \left\{\vphantom{\frac12}\right. 
\frac12 \left(
R^{(+)}_{[\underline{a}}{}^{\underline{egh}}
-{\cal T}_{[\underline{a}}{}^{\underline{egh}}\right)
\left[\vphantom{\frac12}\right. \left(R^{(+)}_{\underline{bc}]\underline{e}}{}^{\underline{f}}
- 2 \omega^{(+)}_{\underline{b}|\underline{ek}|} 
\omega^{(+)}_{\underline{c}}{}^{\underline{fk}} \right)
\omega^{(+)}_{\underline{fgh}} 
\cr 
&&\;\;\;\;\;\;\;\;\;\;\;\;\;\;\;\;\;\;\;\;\;\;\;\;\;\;\;\;\;\;\;\;\;\;\;\;\;\;\;\;\;\;\;\;\; -\ 2  \left(R^{(+)}_{\underline{bc}]\underline{g}}{}^{\underline{f}}
+\ 2 \omega^{(+)}_{\underline{b}|\underline{gk}|} 
\omega^{(+)}_{\underline{eh}}{}^{\underline{f}} \right) \omega^{(+)}_{\underline{ehf}} 
\left.\vphantom{\frac12}\right]\cr
&& + \left( \frac12 
R^{(+)}_{[\underline{a}}{}^{\underline{egh}}
-{\cal T}_{[\underline{a}}{}^{\underline{egh}}\right)
\left[\vphantom{\frac12}\right. 
2 \hat{D}^{(+)}_{\underline{b}} R^{(+)}_{\underline{c}]\underline{egh}}
-  R^{(+)}_{\underline{bc}]\underline{ef}} 
\omega^{(+)}{}^{\underline{f}}{}_{\underline{gh}}
+ 2\  R^{(+)}_{\underline{bc}]\underline{gf}} 
\omega^{(+)}{}^{\underline{eh}}{}_{\underline{f}}
\cr
&&\;\;\;\;\;\;\;\;\;\;\;\;\;\;\;
+ 2\ \left( \frac12 
R^{(+)}_{\underline{b}|\underline{fgh}|}
-{\cal T}_{\underline{b}|\underline{fgh}|}\right)
\omega^{(+)}_{\underline{c}]\underline{e}}{}^{\underline{f}}
-  2\ \left( \frac12 
R^{(+)}_{\underline{b}|\underline{egh}|}
-{\cal T}_{\underline{b}|\underline{egh}|}\right)
\omega^{(+)}_{\underline{c}]\underline{h}}{}^{\underline{f}} 
\left.\vphantom{\frac12}\right]\cr
&&
+\frac12 \omega^{(+)}_{[\underline{a}}{}^{\underline{ef}} \left[\vphantom{\frac12}\right.
2 \hat{D}^{(+)}_{\underline{b}}R^{(+)}_{\underline{c}]\underline{e}}{}^{\underline{gh}} 
\omega^{(+)}_{\underline{fgh}} 
- R^{(+)}_{\underline{b}|\underline{e}|}{}^{\underline{gh}} 
R^{(+)}_{\underline{c}] \underline{fgh}} 
-\frac12 R^{(+)}_{\underline{bc}]\underline{e}}{}^{\underline{k}}
\omega^{(+)}_{\underline{k}}{}^{\underline{gh}} 
\omega^{(+)}_{\underline{fgh}}
+ R^{(+)}_{\underline{bc}]}{}^{\underline{gk}]}
\omega^{(+)}_{\underline{e}}{}^{\underline{h}}{}_{\underline{k}}
\omega^{(+)}_{\underline{fgk}}
\left.\vphantom{\frac12}\right]
\cr
&&
+\omega^{(+)}_{[\underline{a}}{}^{\underline{ef}} \left[\vphantom{\frac12}\right.
2 \hat{D}^{(+)}_{\underline{b}}R^{(+)}_{\underline{c}]\underline{hek}}
\omega^{(+)}{}^{\underline{h}}{}_{\underline{f}}{}^{\underline{k}} 
- R^{(+)}_{\underline{b}|\underline{hek}|}
R^{(+)}_{\underline{c}]}{}^{\underline{h}}{}_{\underline{f}}{}^{\underline{k}} 
-\frac12 R^{(+)}_{\underline{bc}]\underline{h}}{}^{\underline{g}}
\omega^{(+)}_{\underline{gek}}
\omega^{(+)}{}^{\underline{h}}{}_{\underline{f}}{}^{\underline{k}} \cr
&&
\;\;\;\;\;\;\;\;\;\;\;\;\;\;\;\;\;\;\;\;\;\;\;\;\;\;\;
+ \frac12 R^{(+)}_{\underline{bc}]\underline{e}}{}^{\underline{g}]}
\omega^{(+)}_{\underline{hkg}}
\omega^{(+)}{}^{\underline{h}}{}_{\underline{f}}{}^{\underline{k}}
- \frac12 R^{(+)}_{\underline{bc}]\underline{k}}{}^{\underline{g}]}
\omega^{(+)}_{\underline{heg}}
\omega^{(+)}{}^{\underline{h}}{}_{\underline{f}}{}^{\underline{k}}
\left.\vphantom{\frac12}\right]\;,
\end{eqnarray}
where
\begin{eqnarray}
{\cal T}_{\underline{abcd}}&:=&\delta_{\underline{cd}}^{\overline{ef}} 
{\cal T}_{\underline{ab}\overline{ef}} \cr
&=& \frac12 \left(D_{\underline{a}} \omega^{(+)}_{\underline{bcd}} + \omega_{\underline{ab}}{}^{\underline{k}} \omega^{(+)}_{\underline{kcd}} 
- 2 \omega^{(+)}_{\underline{a}[\underline{c}}{}^{\underline{k}} 
\omega^{(+)}_{|\underline{b}|\underline{d}]\underline{k}}\right)\; .
\end{eqnarray}

Then, it introduces extra couplings at ${\cal O}(\alpha'{}^2)$ through $-\frac{1}{12} \hat{H}^2$. 

Because $\hat\Omega^{(2)}$ does not appear to be just an exact form, it cannot be straightforwardly absorbed via a field redefinition of the Kalb–Ramond field. In previous work \cite{Baron:2020xel}, we explicitly verified that a field redefinition of $b_{mn}$ truncates the explicit ${\cal O}(\alpha'{}^2)$ Green–Schwarz transformations,\footnote{Importantly, the leading contribution in the field redefinition in $loc.cit.$ was order $\alpha'{}^2$, hence inducing only an exterior derivative on $H$, at this order.} in agreement with the definition (\ref{tildeH}).  This result is also consistent with the recent article \cite{Gitsis:2026ufz}. 

This leaves us with two logical possibilities: either the $\hat\Omega^{(2)}$ and higher explicit contributions are cancelled by field redefinitions, Bianchi identities, or other manipulations, or there exist two distinct Lorentz-invariant three-forms that differ precisely at ${\cal O}(\alpha'{}^2)$. We hope to return to this issue in a future study.

\section{Conclusions and outlook}
In this work, we have presented a reformulation of the generalized Bergshoeff-de Roo (gBdR) identification that operates directly within the standard supergravity (SUGRA) framework. This marks a significant departure from the original implementations, which were formulated in the T-duality covariant language of Double Field Theory (DFT) and necessitated intricate field redefinitions to connect with the physical Lorentz-covariant degrees of freedom.

Our approach is built upon a specific parametrization of the $O(D,D+k)$ generalized vielbein, combined with a set of gauge-fixing conditions that break the extended Lorentz group to the standard one from the outset. This follows directly from the fact that the Green-Schwarz transformations, generated by identifying the auxiliary gauge group with the Lorentz group, affect the 2-form $\hat{b}_{mn}$ but leave its field strength $\hat{H}_{mnp} = d\hat{b} - \hat{\Omega}_{CS}$ invariant. Consequently, the output of our procedure is naturally expressed in terms of the physical SUGRA fields, thus avoiding the extensive and rapidly growing field redefinitions inherent to the DFT-based approach.

This conceptual simplification translates into a dramatic improvement in computational efficiency. By working directly in the SUGRA frame, we express the generalized Riemann tensor and the resulting action in an extremely compact form. For instance, the six-derivative (order $\alpha'^2$) corrections to the heterotic string action are summarized by only three independent couplings in our formulation, as shown in Eq.~(\ref{Lalpha2}). This is orders of magnitude smaller than the result obtained from the DFT-based gBdR identification.

As a partial check of our construction, we have verified that the pure gravitational couplings at order $\alpha'^2$ are precisely the squared Lorentz Chern-Simons term arising from $\hat{H}^2$, and that all potential $\text{Riemann}^3$ couplings cancel after integration by parts and field redefinitions. This is fully consistent with expectations from the literature.

Looking ahead, our method opens several promising avenues for future research:

Our current analysis has focused on the heterotic string case ($a=0, b\neq 0$). A natural next step is to generalize our parametrization and gauge-fixing to reproduce the full $(a,b)$-family of theories, which includes the bosonic string and the HSZ theory as particular cases.

The approach presented here can be further extended to obtain the $\mathcal{O}(\alpha'^3)$ corrections, which are expected to reproduce the non-$\zeta(3)$ couplings at this order.

The compact form of our results will greatly facilitate a direct comparison with known string scattering amplitude results at order $\alpha'^2$ and beyond. This will allow us to verify the precise mapping between our parameters and the standard $\alpha'$ expansion, and to unambiguously identify any potential scheme-dependent terms.

As we commented in the previous section, there are still some points that deserve attention regarding the derivative corrections of the 3-form at $\mathcal{O}(\alpha'^2)$. These issues should be clarified before extending the analysis to $\mathcal{O}(\alpha'^3)$ couplings.

Finally, it would certainly be interesting to explore the connection between the program introduced here and the one recently developed in the so-called $\alpha'$-bootstrap program by Gitsis, Hassler, and Scala~\cite{Gitsis:2026ufz}.

\section*{Acknowledgments} 
We warmly thank Diego Marques for his comments on the manuscript. Our work is supported by Consejo Nacional de Investigaciones Científicas y Técnicas (CONICET) through the Grant PIP-11220200100981CO and Universidad de La Plata (UNLP).

\begin{appendix}
\section{General parametrization of Coset representatives}

We discuss different parameterizations of the DFT frame, focusing on the construction of a general coset representative. Starting from a representative with partially fixed Lorentz symmetry, the general ungauged parametrization can be restored by applying an arbitrary Lorentz transformation.

As a warm-up, we first consider the $2D$ case, before proceeding to the more general gauged DFT in $2D+k$ dimensions.

\subsection{Generalized frame in the double space}
Consider a vielbein in the triangular gauge for the coset space $O(D,D)/O(D-1,1)\times O(1,D-1)$ 
\begin{eqnarray}
\tilde{E}_{M}{}^{A}=
\frac{1}{\sqrt2}\left(
\begin{matrix}
e^{m}{}_{a} & b_{mp} e^{p}{}_{a}\cr
0 &  e_{m}{}^{a} \end{matrix}
\right)\;,
\end{eqnarray}
where the flat invariant metric in this basis is
\begin{eqnarray}
\tilde{\eta}^{AB}=
\left(
\begin{matrix}
0 & \delta_{a}{}^{b}\cr
\delta^{a}{}_{b} &  0 \end{matrix}
\right)\;.
\end{eqnarray}
Performing a change of basis (acting on flat indices), we obtain a new metric:
\begin{eqnarray}
\eta^{AB}=
\left(
\begin{matrix}
    -g^{\underline{a}\underline{b}} & 0\cr
    0 &  g^{\overline{a}\overline{b}} \end{matrix}
\right)\;.
\end{eqnarray}
In this basis, the vielbein takes the form 
\begin{eqnarray}
\tilde{E}_{M}{}^{A}=
\frac{1}{\sqrt2}\left(
\begin{matrix}
\tilde{e}^{m}{}^{\underline a} & \tilde{e}^{m}{}^{\overline a}\cr
(\bar{b}_{mn} - \bar{g}_{mn})\tilde{e}^{n \underline a} & 
(\bar{b}_{mn} + \bar{g}_{mn})\tilde{e}^{n \overline a} \end{matrix}
\right)\;,
\end{eqnarray}
where $\tilde{e}^{m}{}^{\underline a}= e^{m}{}^{b} \delta_{b}{}^{\underline a}$, $\tilde{e}^{m}{}^{\overline a}= e^{m}{}^{b} \delta_{b}{}^{\overline a}$ are numerically identical vielbeins. This structure corresponds to a partially fixed Lorentz symmetry and is preserved only under Lorentz transformations of the form:
\begin{eqnarray}
O^{A}{}_{B}=(e^{\Lambda})^{A}{}_{B}=
\left(
\begin{matrix}
    O^{\underline a}{}_{\underline b} & 0\cr
    0 &  O^{\overline a}{}_{\overline b} \end{matrix}
\right)\;,
\end{eqnarray}
with $O^{\underline a}{}_{\underline b}= O^{c}{}_{d} \delta_{c}{}^{\underline a} \delta^{d}{}_{\underline b}$ and $O^{\overline a}{}_{\overline b}= O^{c}{}_{d} \delta_{c}{}^{\overline a} \delta^{d}{}_{\overline b}$ representing the same Lorentz transformation. This implies that the double Lorentz group is broken to the ordinary Lorentz group $O(D-1,1)$ determined by $O^{c}{}_{d}$.

To restore the full double Lorentz symmetry, we act on $\tilde E_{M}{}^{\cal A}$ with an arbitrary element $O=e^{\Lambda}$, where
\begin{eqnarray}
\Lambda^{A}{}_{B}=
\left(
\begin{matrix}
    \Lambda^{\underline a}{}_{\underline b} & 0\cr
    0 &  \Lambda^{\overline a}{}_{\overline b} \end{matrix}
\right)\;,
\end{eqnarray}
with $\Lambda^{\underline a}{}_{\underline b}= \underline{\Lambda}^{c}{}_{d} \delta_{c}{}^{\underline a} \delta^{d}{}_{\underline b}$ and $\Lambda^{\overline a}{}_{\overline b}= \overline{\Lambda}^{c}{}_{d} \delta_{c}{}^{\overline a} \delta^{d}{}_{\overline b}$ where $\underline{\Lambda}$ and $\overline{\Lambda}$ are independent transformations. The resulting vielbein is
\begin{eqnarray}
E_{M}{}^{A}=
\frac{1}{\sqrt2}\left(
\begin{matrix}
\bar{e}^{m}{}^{\underline a} & \bar{e}^{m}{}^{\overline a}\cr
(\bar{b}_{mn} - \bar{g}_{mn})\bar{e}^{n \underline a} & 
(\bar{b}_{mn} + \bar{g}_{mn})\bar{e}^{n \overline a} \end{matrix}
\right)\;,
\end{eqnarray}
where now $\overline{e}^{m}{}_{\underline a}=e^{m}{}_{b}\underline{O}^{b}{}_{c}\delta^{c}_{\underline{a}}$ and $\overline{e}^{m}{}_{\overline a}=e^{m}{}_{b}\overline{O}^{b}{}_{c}\delta^{c}_{\overline{a}}$ are two distinct vielbeins of the same metric.  

\subsection{Generalized frame in the extended space}

The same construction applies to the generalized vielbein in the extended $2D+k$ space. Starting from a particular coset representative
\begin{eqnarray}
\tilde{\cal E}_{\cal M}{}^{\cal A}=
\frac{1}{\sqrt2}\left(
\begin{matrix}
e^{m}{}_{a} & 0 & 0\cr
- e^{p}{}_{a} \ c_{pm}  &  e_{m}{}^{a} & A_{m}{}^{\overline{\alpha}}\cr
- A_{p\tilde\mu} e^{p}{}_{a} & 0 & e_{\tilde\mu}{}^{\overline \alpha}
\end{matrix}
\right)\;,
\end{eqnarray}
with $A_{p\tilde\mu}:=e_{\tilde\mu}{}^{\overline\alpha} A_{p \overline{\alpha}}$ and $c_{mn}=b_{mn}+\frac12 A_{m}{}^{\tilde\mu} A_{n\tilde\mu}=b_{mn}+\frac12 A_{m}{}^{\overline{\alpha}} A_{n\overline{\alpha}}$. Or equivalently 
\begin{eqnarray}
{\cal E}_{\cal M}{}^{\cal A}=
\frac{1}{\sqrt2}\left(
\begin{matrix}
\tilde{e}^{m}{}^{\underline a} & \tilde{e}^{m}{}^{\overline a} & 0 \cr
- (\bar{c}_{pm} + \bar{g}_{pm})\tilde{e}^{p \underline a} & 
-(\bar{c}_{pn} + \bar{g}_{pn})\tilde{e}^{p \overline a} & \sqrt{2}\ A_{m}{}^{\overline{\alpha}} \cr
- A_{p\tilde\mu} \tilde{e}^{p \underline{a}} & A_{p\tilde\mu} \tilde{e}^{p\overline{a}} & \tilde{e}_{\tilde\mu}{}^{\overline{\alpha}}
\end{matrix}
\right)\;,\label{gaugedE}
\end{eqnarray}
where $\tilde{e}^{m}{}^{\underline a}= e^{m}{}^{b} \delta_{b}{}^{\underline a}$, $\tilde{e}^{m}{}^{\overline a}= e^{m}{}^{b} \delta_{b}{}^{\overline a}$ are numerically the same vielbein. 

These vielbeins correspond to the same object in different bases, having the following flat invariant metrics
\begin{eqnarray}
\tilde{\eta}^{\cal AB}=
\left(
\begin{matrix}
0 & \delta_{a}{}^{b} & 0\cr
\delta^{a}{}_{b} &  0 &0 \cr
0 & 0 & \kappa^{\overline{\alpha}\overline{\beta}}\end{matrix}
\right)\; ,\;{\rm and} \;\;\;\;\;\;\;\;
\eta^{\cal AB}=
\left(
\begin{matrix}
    -g^{\underline{a}\underline{b}} & 0 & 0\cr
    0 &  g^{\overline{a}\overline{b}} & 0 \cr
    0 & 0 & \kappa^{\overline{\alpha}\overline{\beta}}
    \end{matrix}
\right)\;,
\end{eqnarray}
respectively. 

This parametrization again breaks the extended Lorentz group. Partial restoration is achieved by acting with $O=e^\Gamma$, where
\begin{eqnarray}
\Gamma^{\cal A}{}_{\cal B}=
\left(
\begin{matrix}
    \Gamma^{\underline a}{}_{\underline b} & 0 & 0 \cr
    0 &  \Gamma^{\overline a}{}_{\overline b} & 0 \cr
    0 & 0 & \Gamma^{\overline \alpha}{}_{\overline{\beta}} 
    \end{matrix}
\right)\;,
\end{eqnarray}
resulting in a generalized frame of the same form but with $\tilde{e}^{m\underline{a}}\to \bar{e}^{m\underline{a}}$, and $\tilde{e}^{m\overline{a}}\to \hat{e}^{m\overline{a}}$, now two independent vielbeins of the same metric (i.e. $\bar{e}^{m\underline a} g_{\underline{ab}} \bar{e}^{n\underline{b}} = \bar{e}^{m\overline a} g_{\overline{ab}} \bar{e}^{n\overline{b}}$). 

Full restoration of the extended Lorentz symmetry requires instead:
\begin{eqnarray}
\Gamma^{\cal A}{}_{\cal B}=
\left(
\begin{matrix}
    \Gamma^{\underline a}{}_{\underline b} & 0  \cr
    0 & \Gamma^{\overline{\cal A}}{}_{\overline{\cal B}}  
    \end{matrix}
\right)\;,\;\;\;\;\;\; {\rm with} \;\;\;\;\;
\Gamma^{\overline{\cal A}}{}_{\overline{\cal B}}=
\left(
\begin{matrix}
 \Gamma^{\overline a}{}_{\overline b} & \Gamma^{\overline a}{}_{\overline \beta} \cr
 \Gamma^{\overline \alpha}{}_{\overline b} & \Gamma^{\overline \alpha}{}_{\overline{\beta}} 
\end{matrix}
\right)\;.
\end{eqnarray}
This yields the general parametrization (\ref{ParHet}), which we display here for completeness
\begin{eqnarray}
{\cal E}_{{\cal M}}{}^{\cal A}=
\left(
\begin{matrix}
\frac{1}{\sqrt 2} \hat{e}^{m \underline{a}} & 
\frac{1}{\sqrt 2} \hat{e}^{m \overline{a}} &
\frac{1}{\sqrt 2} \hat{e}^{m \overline{\alpha}} \cr
-\frac{1}{\sqrt 2} \left(\hat{c}_{pm}+ \hat{g}_{pm}\right)\hat{e}^{p \underline{a}} & 
- \frac{1}{\sqrt 2} \left(\hat{c}_{pm} - \hat{g}_{pm}\right)\hat{e}^{p \overline{a}} + \hat{A}_{m}{}^{\overline{a}}&
- \frac{1}{\sqrt 2} \left(\hat{c}_{pm} - \hat{g}_{pm}\right)\hat{e}^{p \overline{\alpha}} + \hat{A}_{m}{}^{\overline{\alpha}}\cr
- \frac{1}{\sqrt 2}\hat{A}_{p\tilde\mu} \hat{e}^{p \underline{a}} & 
- \frac{1}{\sqrt 2} \hat{A}_{p\tilde\mu}\hat{e}^{p\overline{a}} + \hat{e}_{\tilde\mu}{}^{\overline{a}}&
- \frac{1}{\sqrt 2} \hat A_{p\tilde\mu} \hat{e}^{p\overline{\alpha}}+ \hat{e}_{\tilde\mu}{}^{\overline{\alpha}}
\end{matrix}
\right)\;,
\end{eqnarray}

with
\begin{eqnarray}
\hat{e}^{m\underline{a}} = \tilde{e}^{m \underline b} O_{\underline b}{}^{\underline{a}} \;,\;\;\;\;
\hat{e}^{m\overline{a}} = \tilde{e}^{m \overline b} O_{\overline b}{}^{\overline{a}} \;,\;\;\;\;
\hat{e}^{m\underline{\alpha}} = \tilde{e}^{m \overline b} O_{\overline b}{}^{\underline{\alpha}} \;,\;\;\;\;
\hat{e}_{\tilde\mu}{}^{\underline{\alpha}} = \tilde{e}^{m \overline b} O_{\overline{\beta}}{}^{\underline{\alpha}} \cr
\hat{A}_{m}{}^{\overline{\alpha}} = A_{m}{}^{\overline{\beta}} O_{\overline{\beta}}{}^{\overline{\alpha}} \;,\;\;\;\;
\hat{A}_{m}{}^{\overline{a}} = A_{m}{}^{\overline{\beta}} O_{\overline{\beta}}{}^{\overline{a}}\;.
\;\;\;\;\;\;\;\;\;\;\;\;\;\;\;\;\;\;\;\;\;\;\;\;\;\;\;
\end{eqnarray}

The field $e^{m\underline a}$ still describes the original metric $\hat{g}$:
\begin{eqnarray}
\hat{g}^{mn} := \hat{e}^{m\underline a} g_{\underline{ab}} \hat{e}^{n\underline b} =  \tilde{e}^{m\underline a}  \ g_{\underline{ab}} \ \tilde{e}^{n\underline d}     \;,
\end{eqnarray}
while $\hat{e}^{m\overline {a}}$ is no longer identified as the vielbein: $\hat{g}^{mn}\neq \hat{e}^{m\overline a} g_{\overline{ab}} \hat{e}^{n\overline b}$. Instead, 
\begin{eqnarray}
\hat{g}^{mn} &:=& \hat{e}^{m\overline a} g_{\overline{ab}} \hat{e}^{n\overline b} 
+ \hat{e}^{m\overline \alpha} \kappa_{\overline{\alpha\beta}} \hat{e}^{n\overline \beta} \cr
&=& \tilde{e}^{m\overline c} \ O_{\overline{c}}{}^{\overline{A}} \ \eta_{\overline{AB}} \ O_{\overline{d}}{}^{\underline{B}} \ \hat{e}^{n\underline d} \cr
&=& \tilde{e}^{m\overline c}  \ g_{\overline{cd}} \ \tilde{e}^{n\overline d} \;.    
\end{eqnarray}
Additionally, the fields in the parametrization (\ref{ParHet}) obey constraints: (\ref{Id1})-(\ref{Id4}). For instance,
\begin{eqnarray}
\hat A_{m \overline{\cal A}} \ \hat e_{n}{}^{ \overline{\cal A}} = \hat A_{m \overline{\alpha}} O^{\overline{\beta}}{}_{\overline{\cal A}} O_{\overline{c}}{}^{\overline{\cal A}} \ \hat e_{n}{}^{ \overline{c}}=0 \;,  
\end{eqnarray}
due to orthogonality $O^{\overline{\beta}}{}_{\overline{\cal A}} \ O_{\overline{c}}{}^{\overline{\cal A}}=\delta^{\overline{\beta}}{}_{\overline{c}}=0$.

\section{Auxiliary computations}\label{AppB}
We present here some intermediate calculations for those components of the generalized fluxes (\ref{ExtendedFluxes}) being relevant for this work: ${\cal F}_{\underline{a}\overline{\cal CD}}$.

We assume here the gauge fixing conditions to hold. In particular $\hat{f}_{\cal MN}{}^{\cal P}\to g \ f_{\tilde\mu\tilde\nu}{}^{\tilde{\rho}}$ and ${\cal E}^{\tilde\mu}{}_{\overline{a}}=0$ implies that ${\hat f}_{\cal M N P} {\cal E}^{\cal M}{}_{\cal A} {\cal E}^{\cal N}{}_{\cal B} {\cal E}^{\cal P}{}_{\cal C}$ contributes for ${\cal F}_{\underline{a}\overline{\beta\gamma}}$, but do not contribute for  ${\cal F}_{\underline{a}\overline{b\gamma}}$ and  ${\cal F}_{\underline{a}\overline{bc}}$.

Starting with the computation of the generalized Weitzenbock connection components $\Omega_{\underline a \overline{bc}}$ and $\Omega_{\overline{bc} \underline a}$ with the previous gauge fixing conditions and the strong constraint manifestly solved: $\partial^{m}=0,\partial^{\tilde\nu}=0$.
\newpage

\begin{eqnarray}
  \Omega_{\underline a \overline{bc}}&=&
 {\cal E}^{\cal M}{}_{\underline a} \left( \partial_{\cal M}  {\cal E}^{\cal N}{}_{\overline b} \right)  {\cal E}_{{\cal N}}{}_{\overline c}  \cr
 &=&  {\cal E}^{m}{}_{\underline a} \left( \partial_{m}  {\cal E}^{n}{}_{\overline b} \right) {\cal E}_{n}{}_{\overline c}
 +  {\cal E}^{m}{}_{\underline a} \left( \partial_{m}  {\cal E}_{n}{}_{\overline b} \right)  {\cal E}^{n}{}_{\overline c}\cr
 &=& -\frac{1}{(\sqrt2)^3} \hat{e}^{m}{}_{\underline a} \ \hat{e}^{n}{}_{\underline e} \ \hat{e}^{p}{}_{\underline g} \ \partial_{m}\hat b_{np} \  (\chi^{-\frac12})_{\underline{d}}{}^{\underline{e}} \ 
 (\chi^{-\frac12})_{\underline{f}}{}^{\underline{g}} \ \delta^{\underline{fd}}_{\overline{bc}}\cr
 && -\frac{1}{\sqrt2} \hat{e}^{m}{}_{\underline a} \left( \partial_{m} (\chi^{-\frac12})_{\underline{d}}{}^{\underline{e}} \right) 
 (\chi^{\frac12})_{\underline{f}}{}_{\underline{e}} \ \delta^{\underline{fd}}_{\overline{bc}}\cr
 && +\frac{1}{\sqrt2} W_{\underline{aeg}}  (\chi^{\frac12})_{\underline{f}}{}^{\underline{g}} 
 (\chi^{-\frac12})_{\underline{d}}{}^{\underline{e}} \ \delta^{\underline{fd}}_{\overline{bc}}\ ,
\end{eqnarray}
where
$W_{\underline{aeg}}=\hat{e}^{m}{}_{\underline a} \left(\partial_{m} \hat{e}^{n}{}_{\underline e}\right) \hat{e}_{n}{}_{\underline g}$ in the last line is the ordinary Weitzenbock connection. 

Similarly, we find
\begin{eqnarray}
 \Omega_{ \overline{bc} \underline a}&=&
   {\cal E}^{m}{}_{\overline b} \left( \partial_{m}  {\cal E}^{n}{}_{\overline c} \right)  {\cal E}_{n}{}_{\underline a}
 +  {\cal E}^{m}{}_{\overline b} \left( \partial_{m} {\cal E}_{n}{}_{\overline c} \right)  {\cal E}^{n}{}_{\underline a}\cr
 &=& -\frac{1}{(\sqrt2)^3} \hat{e}^{m}{}_{\underline a} \ \hat{e}^{n}{}_{\underline e} \ \hat{e}^{p}{}_{\underline g} \ \partial_{p}\hat b_{mn} \  (\chi^{-\frac12})_{\underline{d}}{}^{\underline{e}} \ 
 (\chi^{-\frac12})_{\underline{f}}{}^{\underline{g}} \ \delta^{\underline{f}}_{\overline{b}} \delta^{\underline{d}}_{\overline{c}}\cr
 && + \frac{1}{(\sqrt2)^3} \hat{e}^{m}{}_{\underline e} \left( \partial_{m} \chi_{\underline{ag}}\right) 
 (\chi^{-\frac12})_{\underline{d}}{}^{\underline{e}}
 (\chi^{-\frac12})_{\underline{f}}{}^{\underline{g}} \ \delta^{\underline{f}}_{\overline{c}} \delta^{\underline{d}}_{\overline{b}}\cr
 && +\frac{1}{(\sqrt2)^3}  W_{\underline{egh}}  (\chi^{-\frac12})_{\underline{d}}{}^{\underline{e}} 
 (\chi^{-\frac12})_{\underline{f}}{}^{\underline{g}} 
 \chi_{\underline a}{}^{\underline h} \ \delta^{\underline{f}}_{\overline{c}} \delta^{\underline{d}}_{\overline{b}}\cr
 && +\frac{1}{(\sqrt2)^3}  W_{\underline{eag}}  (\chi^{-\frac12})_{\underline{d}}{}^{\underline{e}} 
 (\chi^{\frac12})_{\underline{f}}{}^{\underline{g}} 
 \  \delta^{\underline{f}}_{\overline{c}} \delta^{\underline{d}}_{\overline{b}}\ .
\end{eqnarray}
Hence
\begin{eqnarray}
{\cal F}_{\underline a \overline{bc}} &=& 
{\Omega}_{\underline{a} \overline{bc}} 
+ 2 {\Omega}_{[\overline{bc}]\underline{a}}\cr
&=& \frac{1}{\sqrt{2}} \delta^{\underline{fd}}_{\overline{bc}} (\chi^{-\frac12})_{\underline d}{}^{\underline e} \ (\chi^{-\frac12})_{\underline f}{}^{\underline g}\left(\vphantom{\frac12}\right.
- \frac12 H_{\underline{aeg}} 
- 2 \ \omega_{[\underline{ae}]\underline{h}}  
\ \chi_{\underline g}{}^{\underline h} 
+  \ \omega_{[\underline{eg}]\underline{h}}  
\ \chi_{\underline a}{}^{\underline h}\cr
&& +  \hat{e}^{m}{}_{\underline a} (\chi^{\frac12})_{\underline{g}}{}^{\underline{h}}\ \partial_{\underline m}  
\ (\chi^{\frac12})_{\underline{he}} 
 -  \hat{e}^{m}{}_{\underline e} \left(\partial_{\underline m}  
\ \chi_{\underline{ag}} \right)  \left.\vphantom{\frac12}\right) \;,
\end{eqnarray}
where we have used $W_{[ab]c}= -\omega_{[ab]c}$ with $\omega_{abc}= e^{m}{}_{a} \omega_{mbc}$ is the flattened spin connection and we have introduced $H_{\underline{abc}} =  \hat{e}^{m}{}_{\underline{a}} \ \hat{e}^{n}{}_{\underline{b}} \ \hat{e}^{p}{}_{\underline{c}} \ 3\ \partial_{[m}\hat{b}_{np]}$. 

It is convenient to write the fluxes directly in terms of the (Lorentz and gauge invariant) curvatures 
\begin{eqnarray}
\hat{H}_{\underline{abc}}= H_{\underline{abc}} - \left[\hat \Omega_{CS}(A)\right]_{\underline{abc}} \ ,    
\end{eqnarray}
where $\hat\Omega_{CS}(A)$ is the Chern-Simons form (\ref{OCS}), which reads, after the generalized BdR identification as
\begin{eqnarray}
 \left[\hat \Omega_{CS}(A)\right]_{\underline{aeg}}&=& 6\ \hat{e}^{n}{}_{[\underline e} \partial_{n} \hat{e}^{p}{}_{\underline g} \hat{e}_p{}^{\underline i} \left(g_{\underline{a}]\underline{i}}+ \chi_{\underline{a}]\underline{i}}\right)   
 + 6\  \hat{e}^{n}{}_{[\underline{e}} 
 \left(\partial_{n}  {\cal E}^{\tilde\mu}{}_{\underline{g}}\right)
  {\cal E}_{\tilde\mu}{}_{\underline{a}]}
 -2\sqrt{2}  \ \hat{f}_{\tilde\mu\tilde\nu\tilde\rho} 
 {\cal E}^{\tilde\mu}{}_{\underline{a}}
  {\cal E}^{\tilde\nu}{}_{\underline{e}}
  {\cal E}^{\tilde\rho}{}_{\underline{g}}\cr
 &=& -6\omega_{[\underline{aeg}]} 
 -6 \omega_{[\underline{ae}}{}^{\underline{h}} \chi_{\underline{g}]\underline{h}}
 + \frac{6}{g^2\ X_R} \hat{e}^{m}{}_{[\underline{a}}\partial_{m} {\cal F}_{\underline{e}}{}^{\overline{\cal CD}}
 \  {\cal F}_{\underline{g}]}{}_{\overline{\cal CD}}
 + \frac{4\sqrt{2}}{g^2 \ X_R} 
  {\cal F}_{\underline{a}}{}_{\overline{\cal A}}{}^{\overline{\cal C}}   
  {\cal F}_{\underline{e}}{}_{\overline{\cal C}}{}_{\overline{\cal B}}
  {\cal F}_{\underline{g}}{}^{\overline{\cal A}}{}^{\overline{\cal B}}\;, \;\;\;\;\;\;\;\;\;\;
\end{eqnarray}
so we get 
\begin{eqnarray}
{\cal F}_{\underline a \overline{bc}} &=& 
 \frac{1}{\sqrt{2}} \delta^{\underline{fd}}_{\overline{bc}} (\chi^{-\frac12})_{\underline d}{}^{\underline e} \ (\chi^{-\frac12})_{\underline f}{}^{\underline g}\left(\vphantom{\frac12}\right.
- \frac12 \hat{H}_{\underline{aeg}}
+ 3\ \omega_{[\underline{aeg}]} 
- 4 \ \omega_{[\underline{ae}]\underline{h}}  
\ \chi_{\underline g}{}^{\underline h} \cr
&& +  \hat{e}^{m}{}_{\underline a} (\chi^{\frac12})_{\underline{g}}{}^{\underline{h}}\ \partial_{\underline m}  
\ (\chi^{\frac12})_{\underline{he}} 
 - \hat{e}^{m}{}_{\underline e} \left(\partial_{\underline m}  
\ \chi_{\underline{ag}} \right)   \cr
&&- \frac{3}{g^2\ X_R} \hat{e}^{m}{}_{[\underline{a}}\partial_{m} {\cal F}_{\underline{e}}{}^{\overline{\cal CD}}
 \  {\cal F}_{\underline{g}]}{}_{\overline{\cal CD}}
 - \frac{2\sqrt{2}}{g^2 \ X_R} 
  {\cal F}_{\underline{a}}{}_{\overline{\cal A}}{}^{\overline{\cal C}}   
  {\cal F}_{\underline{e}}{}_{\overline{\cal C}}{}_{\overline{\cal B}}
  {\cal F}_{\underline{g}}{}^{\overline{\cal A}}{}^{\overline{\cal B}}  \left.\vphantom{\frac12}\right)\cr 
  &=& 
 \frac{1}{\sqrt{2}} \delta^{\underline{fd}}_{\overline{bc}} (\chi^{-\frac12})_{\underline d}{}^{\underline e} \ (\chi^{-\frac12})_{\underline f}{}^{\underline g}\left(\vphantom{\frac12}\right.
- \frac12 \hat{H}_{\underline{aeg}}
+ 3\ \omega_{[\underline{aeg}]} 
- 4 \ \omega_{[\underline{ae}]\underline{h}}  
\ \chi_{\underline g}{}^{\underline h} \cr
&& +  \hat{e}^{m}{}_{\underline a} \left(\partial_{\underline m}  
\ (\chi^{\frac12})_{\underline{he}} \right)
(\chi^{\frac12})_{\underline{g}}{}^{\underline{h}}\
- \frac{1}{g^2\ X_R} \hat{e}^{m}{}_{\underline{a}} \left( \partial_{m} {\cal F}_{\underline{e}}{}^{\overline{\cal CD}} \right)
 \  {\cal F}_{\underline{g}}{}_{\overline{\cal CD}}\cr
&&+ \frac{2}{g^2\ X_R} \hat{e}^{m}{}_{\underline{e}}\partial_{m} {\cal F}_{\underline{a}}{}^{\overline{\cal CD}}
 \  {\cal F}_{\underline{g}}{}_{\overline{\cal CD}}
 - \frac{2\sqrt{2}}{g^2 \ X_R} 
  {\cal F}_{\underline{a}}{}_{\overline{\cal A}}{}^{\overline{\cal C}}   
  {\cal F}_{\underline{e}}{}_{\overline{\cal C}}{}_{\overline{\cal B}}
  {\cal F}_{\underline{g}}{}^{\overline{\cal A}}{}^{\overline{\cal B}}  \left.\vphantom{\frac12}\right)\; . \cr&&
\end{eqnarray}

Notice that ${\cal F}$ appears in ${\cal L}$ with a $\frac{1}{g^2}$ factor and $\hat{e}^{m}{}_{\underline a} \left(\partial_{\underline m}  
\ (\chi^{\frac12})_{\underline{he}} \right)
(\chi^{\frac12})_{\underline{g}}{}^{\underline{h}}$ contribute at order ${\cal O}(\frac{1}{g^4})$ in ${\cal{F}}_{\underline{a}\overline{bc}}$, so this term will be relevant only at order $\alpha'{}^3$ or higher. 

Another alternative expression is
\begin{eqnarray}
{\cal F}_{\underline a \overline{bc}} &=& 
\delta^{\underline{df}}_{\overline{bc}} {\cal F}^{(+)}_{\underline{adf}}\; ,
\end{eqnarray}
with
\begin{eqnarray}
 {\cal F}^{(+)}_{\underline{adf}} &=& 
\frac{1}{\sqrt{2}} (\chi^{-\frac12})_{[\underline d}{}^{\underline e} \ (\chi^{-\frac12})_{\underline f]}{}^{\underline g}\left(\vphantom{\frac12}\right.
 \hat\omega^{(+)}_{\underline{aeg}} 
- \frac{4}{g^2 \ X_R} \ \omega_{[\underline{ae}]\underline{h}} 
{\cal F}^{\underline{h}\overline{\cal CD}}
\ {\cal F}_{\underline{g}\overline{\cal CD}} \cr
&& -  \hat{e}^{m}{}_{\underline a} \left(\partial_{\underline m}  
\ (\chi^{\frac12})_{\underline{he}} \right)
(\chi^{\frac12})_{\underline{g}}{}^{\underline{h}}\
+ \frac{1}{g^2\ X_R} \hat{e}^{m}{}_{\underline{a}} \left( \partial_{m} {\cal F}_{\underline{e}}{}^{\overline{\cal CD}} \right)
 \  {\cal F}_{\underline{g}}{}_{\overline{\cal CD}}\cr
&&- \frac{2}{g^2\ X_R} \hat{e}^{m}{}_{\underline{e}}\partial_{m} {\cal F}_{\underline{a}}{}^{\overline{\cal CD}}
 \  {\cal F}_{\underline{g}}{}_{\overline{\cal CD}}
 + \frac{2\sqrt{2}}{g^2 \ X_R} 
  {\cal F}_{\underline{a}}{}_{\overline{\cal A}}{}^{\overline{\cal C}}   
  {\cal F}_{\underline{e}}{}_{\overline{\cal C}}{}_{\overline{\cal B}}
  {\cal F}_{\underline{g}}{}^{\overline{\cal A}}{}^{\overline{\cal B}}  \left.\vphantom{\frac12}\right)\; .\cr&&\label{F+}
\end{eqnarray}

Let us compute now ${\Omega}_{\underline{a}\overline{\alpha\beta}}$ and
${\Omega}_{\overline{\alpha\beta}\underline{a}}$ in order to obtain an expression for  ${\cal F}_{\underline{a}\overline{\alpha\beta}}$. 

Instead of computing it from (\ref{parametrization}) we find useful to write these in terms of $\Omega_{\underline{a}\overline{bc}}$ and $\Omega_{\overline{bc}\underline{a}}$.

In order to do this we use the relations
\begin{eqnarray}
{\cal E}^{m}{}_{\overline \alpha} &=& 
 \left( {\cal E}_{\tilde\nu}{}^{\underline{b}} \ e^{\tilde\nu}{}_{\overline{\alpha}} \ \delta_{\underline{b}}^{\overline{a}}\right) \ {\cal E}^{m}{}_{\overline a} , ,\cr
{\cal E}_{m}{}_{\overline \alpha} &=& 
\left( {\cal E}_{\tilde\nu}{}^{\underline{b}} \ e^{\tilde\nu}{}_{\overline{\alpha}} \ \delta_{\underline{b}}^{\overline{a}}\right) \ {\cal E}_{m}{}_{\overline a} 
- \sqrt{2} \ \left({\cal E}_{\tilde\nu}{}^{\underline{b}} \ e^{\tilde\nu}{}_{\overline{\alpha}} \ (\chi^{\frac12})_{\underline{b}}{}^{\underline{a}}\right) \ \hat{e}_{m}{}_{\underline a} \; .
\end{eqnarray}
This yields
\begin{eqnarray}
{\Omega}_{\underline{a}\overline{\alpha\beta}}
&=& e^{\tilde\mu}{}_{[\overline{\alpha}} e^{\tilde\nu}{}_{\overline{\beta}]}
\ {\cal E}_{\tilde\mu}{}^{\underline{e}} \ {\cal E}_{\tilde\nu}{}^{\underline{f}} \
\delta^{\overline{bc}}_{\underline{ef}} \ 
\Omega_{\underline{a}\overline{bc}} 
+ \sqrt{2}  e^{\tilde\mu}{}_{[\overline{\alpha}} e^{\tilde\nu}{}_{\overline{\beta}]} 
  \ {\cal E}_{\tilde\mu}{}^{\underline{e}} \ {\cal E}_{\tilde\nu}{}^{\underline{f}} \
(\chi^{-\frac12})_{\underline{f}}{}^{\underline{c}}
\left( W_{\underline{ac}}{}^{\underline{d}}  (\chi^{\frac12})_{\underline{de}} 
- D_{\underline{a}}(\chi^{\frac12})_{\underline{ec}} \right)\cr
&&
- \frac{1}{\sqrt{2}}  e^{\tilde\mu}{}_{[\overline{\alpha}} e^{\tilde\nu}{}_{\overline{\beta}]} 
\left[ 
D_{\underline{a}} {\cal E}_{\tilde\mu}{}^{\underline{e}} 
\ {\cal E}_{\tilde\nu}{}_{\underline{e}} 
+ D_{\underline{a}}(\Box^{\frac12})_{\tilde\mu}{}^{\tilde\rho} 
\ (\Box^{\frac12})_{\tilde\nu \tilde\rho} 
\right]\; ,
\cr
{\Omega}_{\overline{\alpha}\underline{a}\overline{\beta}}
&=&  e^{\tilde\mu}{}_{[\overline{\alpha}} e^{\tilde\nu}{}_{\overline{\beta}]}
\ {\cal E}_{\tilde\mu}{}^{\underline{e}} \ {\cal E}_{\tilde\nu}{}^{\underline{f}} \
\delta^{\overline{bc}}_{\underline{ef}} \ 
\Omega_{\overline{b}\underline{a}\overline{c}} 
- \frac{1}{\sqrt{2}}  e^{\tilde\mu}{}_{\overline{\beta}} e^{\tilde\nu}{}_{\overline{\alpha}} 
 (\chi^{-\frac12})_{\underline{f}}{}^{\underline{c}} 
 \left(
{\cal E}_{\tilde\mu}{}^{\underline{e}} \ {\cal E}_{\tilde\nu}{}^{\underline{f}} \
 W_{\underline{ca}}{}^{\underline{d}} 
 (\chi^{\frac12})_{\underline{ed}}
 - D_{\underline{c}} {\cal E}^{\tilde\rho}{}_{\underline{a}} \
  {\cal E}_{\tilde\nu}{}^{\underline{f}}  (\Box^{\frac12})_{\tilde\rho\tilde\mu}
\right)\; ,\cr&&
\end{eqnarray}
where $D_{\underline{a}}= \hat{e}^{m}{}_{\underline{a}}\partial_{m}$ 
and\footnote{Notice that ${\Omega}$ is antisymmetric in the last two indices ${\Omega}_{\cal ABC}=-{\Omega}_{\cal ACB}$, but this symmetry is not inherited by $W$: $W_{\underline{abc}}\neq - W_{\underline{acb}}$. }
\begin{eqnarray}
 {\cal F}_{\underline a \overline{\alpha\beta}} =
 {\Omega}_{\underline{a}\overline{\alpha\beta}} + 2\ {\Omega}_{[\overline{\alpha\beta}]\underline{a}} + g f_{\tilde\mu\tilde\nu\tilde\rho} 
 {\cal E}^{\tilde\mu}{}_{\underline{a}} {\cal E}^{\tilde\nu}{}_{\overline{\alpha}}
{\cal E}^{\tilde\rho}{}_{\overline{\beta}} 
 \end{eqnarray}
is
\begin{eqnarray}
 {\cal F}_{\underline a \overline{\alpha\beta}} &=& 
  e^{\tilde\mu}{}_{[\overline{\alpha}}
 e^{\tilde\nu}{}_{\overline{\beta}]} \left\{\vphantom{\frac12}\right. g f_{\tilde\rho\lambda\sigma} {\cal E}^{\tilde\rho}{}_{\underline{a}} \ (\Box^{\frac12})^{\lambda}{}_{\tilde\mu} \ (\Box^{\frac12})^{\sigma}{}_{\tilde\nu}\cr 
 &&+   {\cal E}_{\tilde\nu}{}^{\underline{f}} \
\left[ {\cal E}_{\tilde\mu}{}^{\underline{e}} \left(\vphantom{\frac12}
\delta^{\overline{bc}}_{\underline{ef}} \ 
{\cal F}_{\underline{a}\overline{bc}} 
-\sqrt{2}(\chi^{-\frac12})_{\underline{f}}{}^{\underline{c}}
\left(D_{\underline{a}}(\chi^{\frac12})_{\underline{ce}} 
+ 2\omega_{[\underline{ac}]}{}^{\underline{d}} (\chi^{\frac12})_{\underline{de}}\right)\right)
+\sqrt{2} (\chi^{-\frac12})_{\underline{f}}{}^{\underline{c}}
D_{\underline{c}}{\cal E}^{\tilde\rho}{}_{\underline{a}}(\Box^{\frac12})_{\tilde\rho\tilde\mu}
\right]\cr
&&-\frac{1}{\sqrt{2}} \left(
D_{\underline{a}}{\cal E}_{\tilde\mu}{}^{\underline{e}} \ 
{\cal E}_{\tilde\nu}{}_{\underline{e}} 
+ D_{\underline{a}}(\Box^{\frac12})_{\tilde\mu}{}^{\tilde\rho} \
(\Box^{\frac12})_{\tilde\nu\tilde\rho}
 \right) \left.\vphantom{\frac12}\right\}\;.
\end{eqnarray}
Turning to $\mathcal{F}_{\underline{a}\overline{b\gamma}}$, one obtains
\newpage
\begin{eqnarray}
{\Omega}_{ \underline{a}\overline{b\gamma}} 
&=&
\left( {\cal E}_{\tilde\nu}{}^{\underline{d}} \ e^{\tilde\nu}{}_{\overline{\gamma}} \ \delta_{\underline{d}}^{\overline{c}}\right) \ 
{\Omega}_{ \underline{a}\overline{bc}}  +\frac{1}{\sqrt{2}} \left( {\cal E}_{\tilde\nu}{}^{\underline{d}} \ e^{\tilde\nu}{}_{\overline{\gamma}} \ \delta_{\overline{b}}^{\underline{e}}\right) (\chi^{-\frac12})_{\underline{e}}{}^{\underline{f}} \ \left( D_{\underline{a}}(\chi^{\frac12})_{\underline{fd}} - W_{\underline{af}}{}^{\underline{c}} (\chi^{\frac12})_{\underline{cd}} \right),\cr 
{\Omega}_{ \overline{b}\underline{a}\overline{\gamma}} 
&=& \left( {\cal E}_{\tilde\nu}{}^{\underline{d}} \ e^{\tilde\nu}{}_{\overline{\gamma}} \ \delta_{\underline{d}}^{\overline{c}}\right) \
{\Omega}_{ \overline{b}\underline{a}\overline{c}}
- \frac{1}{\sqrt{2}} \left( {\cal E}_{\tilde\nu}{}^{\underline{d}} \ e^{\tilde\nu}{}_{\overline{\gamma}} \ \delta_{\overline{b}}^{\underline{e}}\right) (\chi^{-\frac12})_{\underline{e}}{}^{\underline{f}} 
\ W_{\underline{fa}}{}^{\underline{c}} (\chi^{\frac12})_{\underline{cd}} 
+ \frac{1}{\sqrt{2}}   \ e^{\tilde\nu}{}_{\overline{\gamma}} \ \delta^{\underline{e}}_{\overline{b}} (\chi^{-\frac12})_{\underline{e}}{}^{\underline{f}}  \ D_{\underline{f}}({\cal E}^{\tilde\nu}{}_{\underline{a}}) (\Box^{\frac12})_{\tilde\mu\tilde\nu}
,\cr 
{\Omega}_{ \overline{\gamma}\underline{a}\overline{b}} 
&=& 
\left( {\cal E}_{\tilde\nu}{}^{\underline{d}} \ e^{\tilde\nu}{}_{\overline{\gamma}} \ \delta_{\underline{d}}^{\overline{c}}\right) \
{\Omega}_{ \overline{c}\underline{a}\overline{b}}
\ .
\end{eqnarray}
Therefore
\begin{eqnarray}
{\cal F}_{\underline{a} \overline{b\gamma}} &=& 
\left( {\cal E}_{\tilde\nu}{}^{\underline{d}} \ e^{\tilde\nu}{}_{\overline{\gamma}} \ \delta_{\underline{d}}^{\overline{c}}\right) {\cal F}_{\underline{a}\overline{bc}} 
\ + \ \frac{1}{\sqrt{2}} \left( {\cal E}_{\tilde\nu}{}^{\underline{d}} \ e^{\tilde\nu}{}_{\overline{\gamma}} \ \delta^{\underline{e}}_{\overline{b}}\right) 
(\chi^{-\frac12})_{\underline{e}}{}^{\underline{f}}
\ \left( D_{\underline{a}}(\chi^{\frac12})_{\underline{fd}} + 2 \omega_{[\underline{af}]}{}^{\underline{c}} (\chi^{\frac12})_{\underline{cd}} \right)  \cr
&&-  \frac{1}{\sqrt{2}}   \ e^{\tilde\nu}{}_{\overline{\gamma}} \ \delta^{\underline{e}}_{\overline{b}} (\chi^{-\frac12})_{\underline{e}}{}^{\underline{f}}  \ D_{\underline{f}}({\cal E}^{\tilde\nu}{}_{\underline{a}}) (\Box^{\frac12})_{\tilde\mu\tilde\nu}\; .
\end{eqnarray}

\end{appendix}

\end{document}